\documentclass[preprint,aps,showpacs]{revtex4}

\usepackage{graphicx}% Include figure files
\usepackage{dcolumn}% Align table columns on decimal point
\usepackage{bm}% bold math
\def\ä{\"{a}}
\def\ü{\"{u}}
\def\ö{\"{o}}
\def\Ä{\"{A}}
\def\Ü{\"{U}}
\def\Ö{\"{O}}
\begin{document}

\preprint{APS/123-QED}

\title{A comparison of force sensors for atomic force microscopy based on quartz tuning forks and length extensional resonators}% Force line breaks with \\

\author{Franz J. Giessibl}
\email{franz.giessibl@physik.uni-regensburg.de}

\affiliation{%
Universit\"at Regensburg, Institute of Experimental and Applied
Physics,  Universit\"atsstrasse 31, D-93040 Regensburg, Germany.
}%
\author{Toyoaki Eguchi}
\email{eggy@ncassembly.jst.go.jp}
\affiliation{NAKAJIMA Designer Nanocluster Assembly Project, ERATO, Japan Science and Technology Agency (JST)
3-2-1 Sakato, Takatsu-ku,  Kawasaki 213-0012, Japan
\\
Graduate School of Science and Technology, Keio University
3-14-1 Hiyoshi, Kohoku-ku, Yokohama 223-8522, Japan
}

\author{Florian Pielmeier}

\affiliation{%
Universit\"at Regensburg, Institute of Experimental and Applied
Physics,  Universit\"atsstrasse 31, D-93040 Regensburg, Germany.
}%
\author{Toshu An}

\affiliation{%
Institute for Materials Research, Tohoku University,
2-1-1, Katahira, Aoba-ku, Sendai 980-8577, JAPAN
}%

\author{Yukio Hasegawa}

\affiliation{%
Institute for Solid State  Physics, University of Tokyo 5-1-5,
Kashiwa-no-ha, Kashiwa, Chiba 277-8581 Japan.
}%
%\altaffiliation[Also at ]{}%Lines break automatically or can be forced with \\
%\author{}%

\date{ \today
%e.g. Physical Review B
}% It is always \today, today,

\begin{abstract}
The force sensor is key to the performance of atomic force
microscopy (AFM). Nowadays, most AFMs use micro-machined force
sensors made from silicon, but piezoelectric quartz sensors are
applied at an increasing rate, mainly in vacuum. These self
sensing force sensors allow a relatively easy upgrade of a
scanning tunneling microscope to a combined scanning
tunneling/atomic force microscope. Two fundamentally different
types of quartz sensors have achieved atomic resolution: the \lq
needle sensor\rq{} that is based on a length extensional resonator
and the \lq qPlus sensor\rq{} that is based on a tuning fork.
Here, we calculate and measure the noise characteristics of these
sensors. We find four noise sources: deflection detector noise,
thermal noise, oscillator noise and thermal drift noise.
We calculate the effect of these noise sources as a factor of
sensor stiffness, bandwidth and oscillation amplitude.
We find that for self sensing quartz sensors, the deflection detector noise
is independent of sensor stiffness, while the remaining three noise sources
increase strongly with sensor stiffness. Deflection detector noise increases
with bandwidth to the power of 1.5, while thermal noise and oscillator noise
are proportional to the square root of the bandwith. Thermal drift noise, however,
is inversely proportional to bandwidth.
The first
three noise sources are inversely proportional to amplitude while thermal drift noise is independent of the amplitude.
Thus, we show that the earlier finding that quoted optimal signal-to-noise ratio
for oscillation amplitudes similar to the range of the forces is still correct
when considering all four frequency noise contributions. Finally, we suggest how the signal-to-noise ratio of
the sensors can be further improved, briefly discuss the
challenges of mounting tips and compare the noise performance of
self sensing quartz sensors and optically detected Si cantilevers.
\end{abstract}

\maketitle

\tableofcontents

\section{Introduction}
\lq Atomic Force Microscopy\rq{} (AFM) has been introduced in 1986
by Binnig, Gerber and Quate \cite{Binnig1986}. The large number of
citations (the article is now one of the most highly cited
publications that have appeared in \textit{Physical Review
Letters}) show that AFM is an important scientific tool with
fruitful applications in various fields of science. The key
element of AFM is the force sensor that probes the small forces
that act between a sharp tip and a sample. Simplifying the force
sensor and increasing its force resolution and imaging speed are
therefore important tasks.

\begin{figure}
\begin{center}
\includegraphics[clip=true, width=0.8\textwidth]{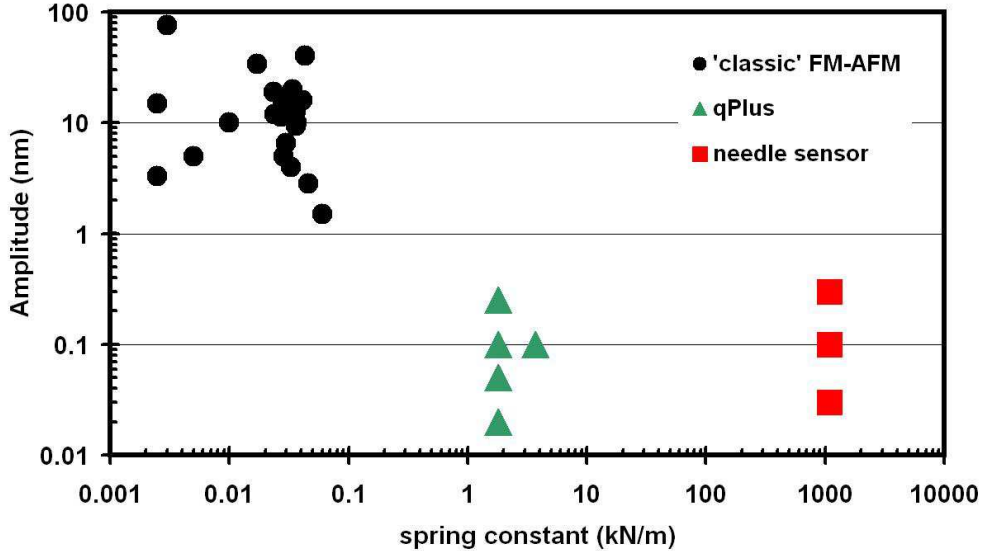}
\end{center}
\caption{(Color online) Parameter fields of cantilever spring
constants $k$ and oscillation amplitudes $A$ for classic Si
cantilevers, qPlus sensors and needle sensors. The $(k,A)$ data
points for Si cantilevers and qPlus sensors are adapted from table
I in \cite{Giessibl2003}, the ones from a shortened qPlus sensors
are taken from \cite{Schmid2008}. To enable stable oscillation of
the cantilever at the optimal amplitudes around 100\,pm, it was
necessary to increase the spring constants of cantilevers (\lq
classic\rq{} FM-AFM) from about 10\,N/m by more than two orders of
magnitude (qPlus sensors). The needle sensor has a stiffness that
is almost three orders of magnitude larger than that of the qPlus
sensor. The question, whether this further increase is beneficial
is addressed in this paper.}\label{fig_k_A_space}
\end{figure}

Atomic resolution by AFM on a reactive surface was first achieved
by frequency modulation AFM (FM-AFM) \cite{Albrecht1991} utilizing
a piezo-resistive silicon cantilever \cite{Tortonese1993} with a
spring constant of $k=17$\,N/m at an oscillation amplitude of
$A=34$\,nm \cite{Giessibl1995}. While atomic resolution on various
surfaces has been obtained with similar combinations of $(k,A)$
(see table I in \cite{Giessibl2003}), a calculation of the
signal-to-noise ratio in FM-AFM as a function of the oscillation
amplitudes yielded an optimal oscillation amplitude that
corresponds to the decay length of the forces that are used for
imaging. The spring constant of the cantilever should be as small
as possible for obtaining a large frequency shift, on the other
hand, the cantilever must be stiff enough to prevent instabilities
such as jump-to-contact \cite{Giessibl1999ASS}. Compared to the
initial parameter set of $(k,A)$ that allowed atomic resolution \cite{Giessibl1995}, the spring constant of
the sensor has to be larger by a factor of about one to two orders of magnitude, and the
amplitude has to be reduced by a factor of two to three orders of magnitude. The
reduced amplitude not only increases the signal-to-noise ratio, it
also reduces the sensitivity to unwanted long-range force
contributions \cite{Giessibl2003}. Figure \ref{fig_k_A_space}
shows the parameters used with \lq classic\rq{} Si cantilevers,
qPlus sensors and needle sensors.

For atomic imaging, it was suggested that the optimal stiffness
$k_{opt}$ is approximately in the interval
\begin{equation}\label{eq_k_opt}
    500\,\textrm{N/m} < k_{opt} < 3000\,\textrm{N/m}
\end{equation}
at amplitudes of
about 100\,pm \cite{Giessibl1999ASS}.

Self-sensing cantilevers such as piezo-resistive silicon
cantilevers or piezo-electric quartz sensors are attractive
because these sensors simply need to be connected to an electronic preamplifier to provide
an electrical deflection signal. In contrast, sensors that utilize deflection
measurements based on electron tunneling \cite{Binnig1986} or
optical means \cite{Meyer1988} require precise mechanical alignment schemes
which can be challenging in vacuum or low temperature environments. Optical
deflection measurements also involve light and heat introduction close to the
sample. For some applications, such as low-temperature measurements or the study
of samples that alter their properties under electromagnetic radiation, optical
deflection measurements are disadvantageous.

Because FM-AFM relies on the
alteration of the oscillation frequency of the cantilever under
the influence of tip-sample force gradients, a high intrinsic
frequency stability of the cantilever is desirable. Silicon
cantilevers, the most widespread type in use, change their
frequency by $-35$\,ppm per K at room temperature
\cite{Gysin2004}. In contrast, quartz resonators change their
frequency by less than 1\,ppm per K as long as their temperature
is kept within $\pm14$\,K of their turnover temperature (see eq.
\ref{eq_temp_drift_f}). The outstanding stability of quartz that
has been utilized since decades for watches and frequency standards provides
for highly stable FM-AFM sensors as well.

Two types of commercially available quartz frequency standards are
particularly well suited for conversion into force sensors: quartz
tuning forks and length extensional resonators (LER).
\begin{figure}
\begin{center}
\includegraphics[clip=true, width=0.8\textwidth]{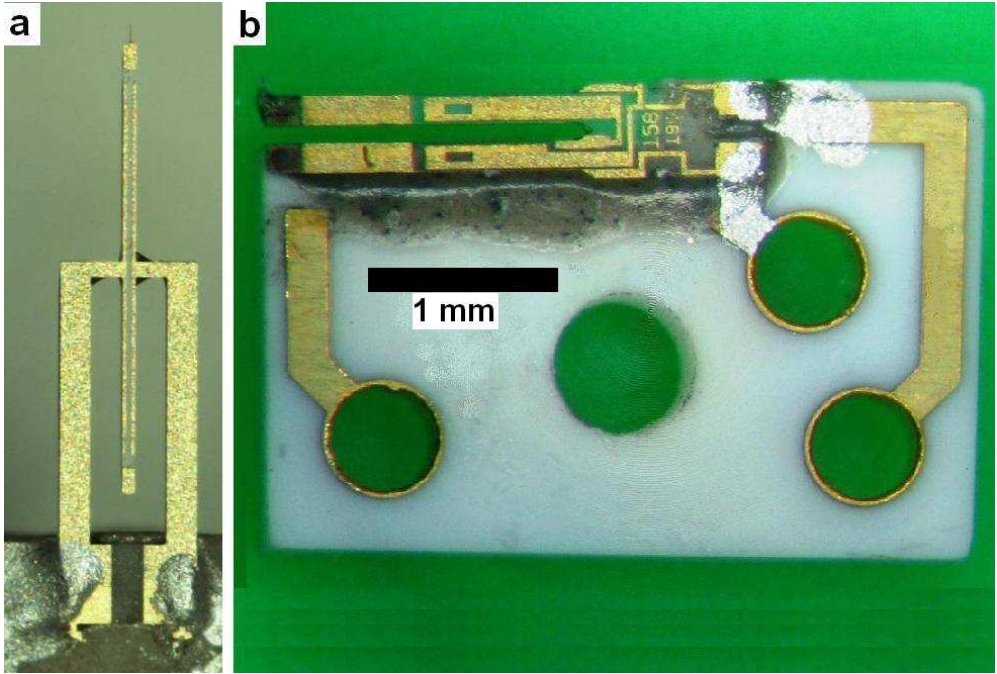}
\end{center}
\caption{(Color online) a) Needle sensor. b) qPlus
sensor. The scale bar is valid for both sensors.}\label{fig_real_sensors}
\end{figure}
Both tuning forks and length extensional resonators essentially
consist of two coupled electromechanical oscillators that have
exactly the same eigenfrequency and oscillate in an antiparallel
mode. Attaching a tip to one of the oscillators changes its
resonance frequency, so the tip either has to be very light or a
similar mass has to be attached to the other oscillator. Force
sensors based on tuning forks have been used by Guethner et al.
\cite{Guethner1989} already in 1989, where a tip has been mounted
onto one prong, and the mass of the tip was balanced with a
counterweight on the other prong \cite{Dransfeld1993}.  The length
extensional resonator has been supplemented by a light tip on one
of its two bars to form the needle sensor after Bartzke et
al.\cite{Bartzke1993,Bartzke1995P} in 1993. The qPlus sensor is
also based on a tuning fork, but one of the prongs is immobilized
by attaching it to a heavy substrate such that the free prong is
essentially a quartz cantilever
\cite{Giessibl1996P,Giessibl1998,Giessibl2000APL}. In this case,
the tip can be massive, and the oscillating tip can interact
vigorously with the sample without a reduction in the $Q$ value.
These sensors with metal probe tips allow a simple implementation
of combined scanning tunneling microscopy (STM) and AFM. Quartz
tuning forks are available with eigenfrequencies $f_0$ ranging
from about 32 to 200\,kHz. Length extensional resonators are
available in eigenfrequencies of 0.5\,MHz to a few MHz \cite{Morawski2011}. In the
comparison here, we focus on a specific type of tuning fork that
is used in SWATCH wristwatches with stiffness $k'=1800$\,N/m and
$f_0=32768$\,Hz and a specific type of length extensional
resonator with $k'=540$\,kN/m and $f_0=1$\,MHz, because these
types were used in the experimental data cited below (see Fig.
\ref{fig_sensorgeom} and Table \ref{table_sensordims} for
geometric details). In section \ref{section_suggestions}, we will
suggest optimized geometries for both types of sensors, but here
we refer to \lq standard qPlus-\rq{} or \lq standard needle
sensor\rq{} as shown in Fig. \ref{fig_real_sensors} to be based on
the geometries as specified in Table \ref{table_sensordims}.

\begin{figure}
\begin{center}
\includegraphics[clip=true, width=0.8\textwidth]{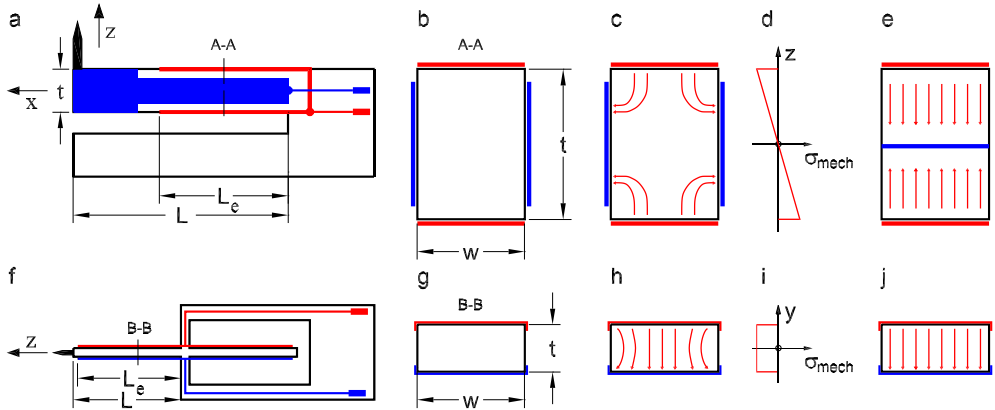}
\end{center}
\caption{(Color online) Geometry of sensors based on quartz tuning forks (a-e)
and length extensional resonators (f-j). A qPlus sensor (a) is
created by attaching one of the prongs
 of the tuning fork to a substrate and attaching a tip to the other prong. For clarity, only the electrodes on the free prong are shown. The prong without
 displayed electrodes is fixed to a massive substrate (not shown here, see Fig. 1 in \cite{Giessibl2000APL}). A needle sensor (f) is built by attaching a light tip to one prong of the length extensional resonator.
 Figures (a,b,f,g) illustrate the geometrical dimensions as listed in table \ref{table_sensordims}, (c,h) show a schematic view of the electrostatic field in the cross sections and (d,i)
 show the mechanical stress profile along a cross section. Figures (e,j) show the idealized field distribution within the quartz crystals.
 The qPlus sensor uses a bending mode, thus the mechanical stress is maximal where the charge-collecting electrodes are located (d), while
 the length extensional resonator develops a uniform stress profile
 (i). The idealized field distribution (e,j) is much closer to the
 actual field distribution (c,h) for the needle sensor (j vs. h) than
 for the qPlus sensor (e vs. c).
}\label{fig_sensorgeom}
\end{figure}
\begin{center}
\begin{table}

\begin{tabular}{|c||c|c|c|c|c|c|c|c|}
    \hline
                    & $L$ ($\mu$m) & $L_{e}$ ($\mu$m) & $t$ ($\mu$m) & $w$ ($\mu$m)& $k'$ (N/m)        & $k$ (N/m)        & $f_0$ (Hz)    \\
    \hline \hline
    needle sensor   & $1340$       &   $1100$         & 70           &     130     & 540\,000          & 1\,080\,000      & 1\,000\,000   \\
    \hline
    qPlus sensor    & $2400$       &   $1600$         & 214          &     126     & 1\,800            & 1\,800           & 32\,768  \\
    \hline
\end{tabular}
\caption{Geometrical parameters, stiffness $k$ and eigenfrequency $f_0$ of the quartz oscillators used. The needle sensor is based on a length extensional
resonator, while the qPlus sensor is based on a quartz tuning fork.}
\end{table} \label{table_sensordims}
\end{center}
%\caption{Dimensions, spring constants and eigenfrequencies of
%qPlus sensors and needle sensors as shown in Fig.
%\ref{fig_sensorgeom}.}

A qPlus sensor with $k=1.8$\,kN/m has allowed subatomic spatial
resolution \cite{Giessibl2000Science,Hembacher2004Science}, atomic
resolution of lateral forces \cite{Giessibl2002PNAS}, simultaneous
force and current spectroscopy on graphite
\cite{Hembacher2005PRL}, the measurement of forces acting in
atomic manipulation \cite{TernesScience2008}, the detection of a
single charge on an atom \cite{GrossScience2009a} and
unprecedented spatial resolution of an organic molecule
\cite{GrossScience2009b} and helped to identify an initially
unidentified organic molecule that was hauled up from the Mariana
Trench \cite{GrossNature2010}. Even more recent, the relationship
between tunneling current and forces has been revealed
\cite{TernesPRL2011} and the interaction of two CO molecules has been studied \cite{SunPRL2011}. Furthermore a reduction of the effective tunneling voltage caused by the flow of current on weakly conductive samples has been detected by a reduced electrostatic attraction \cite{WeymouthPRL2011}.

Although the needle sensor's effective stiffness of more than
1\,MN/m ($k=2k'$, see eq. \ref{eq_keff}) is far beyond the
stiffness range that is suggested to be optimal above, it has
produced atomic resolution on silicon at 4\,K \cite{An2005,An2008}
and at room temperature \cite{Torbrugge2010}. Therefore it is
instructive to analyze the success factors of these sensors for
the purpose of further improving their performance.

\section{Frequency shift as a function of tip-sample interaction for single and coupled oscillators}
\begin{figure}
\begin{center}
\includegraphics[clip=true, width=1\textwidth]{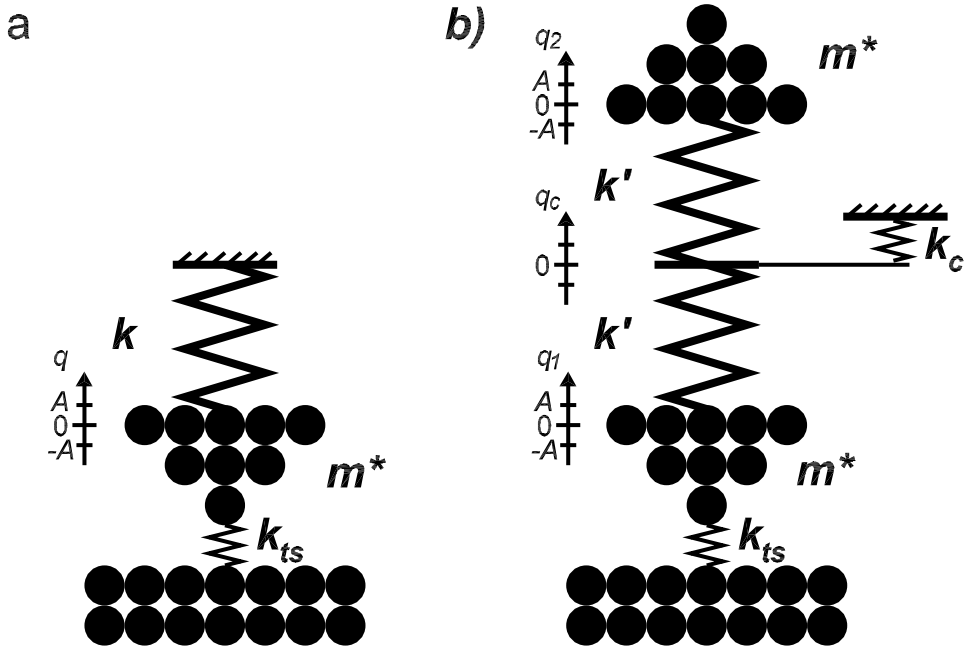}
\end{center}
\caption{a) Mechanical analog of a single oscillator-type force sensor (standard
cantilever or qPlus sensor as in Fig. 1a), consisting of a single oscillating
beam. The single oscillator has only one degree of freedom, its deflection $q$. b) Mechanical analog of a coupled oscillator used as a force sensor (tuning fork
or length extensional resonator as in Fig 1e). The coupled oscillator has three degrees of freedom: the deflection of the central mount $q_c$ and the deflections
of the two coupled oscillators $q_{1,2}$.}
\label{fig_single_coupled_osc}
\end{figure}

In frequency modulation atomic force microscopy, the
eigenfrequency $f$ of a force sensor (such as a qPlus sensor or a needle sensor,
see Fig. \ref{fig_sensorgeom}) that vibrates at a constant
amplitude $A$ changes with the action of force gradients by a frequency shift $\Delta f = f-f_0$.
With $f=f_0+\Delta f$ and $f_{0} = \frac{1}{2\pi}\sqrt{k/m^*}$,
the frequency shift is given by
\begin{equation}
\Delta f =  \frac{f_{0}}{2k}
\langle k_{ts}\rangle\label{gradient_approximation}
\end{equation}
with \cite{Giessibl2001APL}
\begin{equation}
\langle k_{ts}\rangle(z) = \frac{2}{\pi}\int_{-1}^{1}k_{ts}(z+\zeta
A)\sqrt{1-\zeta^2}d\zeta.
\end{equation}
At large amplitudes, the frequency shift is given by
\begin{equation}
\Delta f =
\frac{f_{0}}{k}\frac{1}{A^{3/2}}\gamma_{ts}\label{lA_approximation}
\end{equation}
with the normalized frequency shift $\gamma_{ts} \approx 0.4
F_{ts}\sqrt{\lambda}$ \cite{Giessibl2000}.
When $A$ is very small compared to the decay length $\lambda$ of the
force gradient, $\langle k_{ts}\rangle(z)$ is similar to
$k_{ts}(z)$, the gradient of the tip-sample forces at the center
position of the cantilever that oscillates around $z\pm A$.

The eigenfrequency is found by
solving the equation of motion for the cantilever deflection
$q(t)$, the single degree of freedom:
\begin{equation}
m^*\frac{\partial^2 q}{\partial t^2}  = -q (k+k_{ts})
\end{equation}
resulting in $q(t)=A\cos(\omega t+\phi)$ with
$\omega^2=(k+k_{ts})/m^*$ and $\omega=2\pi f$.

Figure \ref{fig_single_coupled_osc} b) shows a coupled oscillator such as
a tuning fork or a length extensional resonator. In the case of a
coupled oscillator, the oscillator has three degrees of freedom
$q_1(t),\, q_2(t)$ and $q_c(t)$, leading to more complicated modes
than in the case of a cantilever or qPlus sensor with its single
degree of freedom. When the inertial forces (given by mass times
acceleration) of the center piece of the length extensional
resonator (LER) can be neglected (a fair assumption for the antiparallel
mode), the equation of motion is relatively easy to solve:
\begin{eqnarray}
m^*\frac{\partial^2 q_1}{\partial t^2}  &=&  -k_{ts} q_1 +
k' (q_c-q_1)\\
m^*\frac{\partial^2 q_2}{\partial t^2}  &=&  -k' (q_2 - q_c)
\end{eqnarray}
Because the center of the LER needs to be in equilibrium, we find
\begin{equation}
q_c k_c=k'(q_1-q_c)+k'(q_2-q_c)
\end{equation}
With $\kappa=1/(2+k_c/k')$ we can substitute $q_c=\kappa(q_1+q_2)$ and find
\begin{eqnarray}
\frac{\partial^2 q_1}{\partial t^2}  &=&  -\omega_0^2(1+k_{ts}/k'-\kappa) q_1 +
\omega_0^2 \kappa  q_2 \\
\frac{\partial^2 q_2}{\partial t^2}  &=&  +\omega_0^2\kappa q_1 - \omega_0^2(1-\kappa) q_2
\end{eqnarray}
with $\omega_0^2=k'/m^*$. Using a harmonic ansatz $q_{1,2}(t)=A_{1,2}\cos(\omega t+\phi_{1,2})$,
we find two solutions for $\omega$:
\begin{equation}
\omega_{1,2}^2 = \omega_0^2 \{1-\kappa+\frac{k_{ts}}{2k'} \pm
\sqrt{\kappa^2+\frac{k_{ts}^2}{4k'^2}}\}.
\label{eigenvalue_coupled_osc}
\end{equation}
Typically, $\kappa>1/3$ because $k_c<k'$ and with $k_{ts}<<k'$, we
can approximate the square root in eq.
\ref{eigenvalue_coupled_osc}:
\begin{equation}
\omega_{1,2}^2 \approx \omega_0^2 \{1-\kappa+\frac{k_{ts}}{2k'}
\pm \kappa(1+\frac{k_{ts}^2}{8\kappa^2 k'^2})\}.
\label{eigenvalue_coupled_osc2}
\end{equation}

Two solutions are found, where the plus sign in eq.
\ref{eigenvalue_coupled_osc} corresponds to a high-frequency
antiparallel motion ($A_{1} \approx -A_{2},\phi_{1}=\phi_{2}$)
\begin{equation}
\omega_{1}^2 \approx \omega_0^2 \{1+\frac{k_{ts}}{2k'}+
\frac{k_{ts}^2}{8\kappa k'^2}\} \label{eigenvalue_coupled_osc+}
\end{equation}
and the minus sign to a low frequency parallel motion ($A_{1} \approx A_{2},\phi_{1}=\phi_{2}$)
\begin{equation}
\omega_{2}^2 \approx \omega_0^2
\{1-2\kappa+\frac{k_{ts}}{2k'}-\frac{k_{ts}^2}{8 \kappa k'^2}\}.
\label{eigenvalue_coupled_osc-}
\end{equation}

The antiparallel motion is used in force microscopy with coupled
oscillators, where the frequency shift of the sensor is given by
\begin{equation}
\frac{\omega_1-\omega_0}{\omega_{0}} = \frac{\Delta f}{f_{0}} = \frac{k_{ts}}{4 k'}
\label{df_needle}
\end{equation}
(in leading order of $k_{ts}$). The frequency shift for a coupled
oscillator is thus only half the value of the single oscillator
after eq. \ref{gradient_approximation}. We can still use eqs.
\ref{gradient_approximation} and \ref{lA_approximation} by
defining an effective stiffness $k$ that is twice as large as the
individual stiffness $k'$ of each of the two coupled oscillators.
\begin{equation}\label{eq_keff}
k_{coupled}=2k'.
\end{equation}
Equation \ref{gradient_approximation} links the signal (i.e. the
physical observable) to $k_{ts}$, the physical origin of the
signal by multiplying it with the prefactor $f_0/2k$. To obtain a strong signal,
the prefactor $f_0/2k$ should be large. For a tip-sample force gradient of
1\,N/m, a standard needle sensor would yield a frequency shift of
$\Delta f = 0.463\,\textrm{Hz}$, while a standard qPlus
sensor would yield a frequency shift of $\Delta f = 8.33\,\textrm{Hz}$. However, to assess the
signal-to-noise ratio, we need to consider noise as well as signal
strength. Noise also depends on the sensor type and will be
discussed in section \ref{sec_noise}.

\section{Operating Principles and Sensitivity of Quartz Sensors}

\subsection{Sensor based on quartz tuning fork (qPlus sensor)}

For a rectangular cantilever with width $w$, thickness $t$ and
length $L$, the spring constant $k$ is given by \cite{Chen}:
\begin{equation}
k=\frac{Ewt^{3}}{4L^{3}}.\label{k_cl}
\end{equation}
where $E$ is Young's modulus. The fundamental eigenfrequency
$f_{0}$ is given by \citep{Chen}:
\begin{equation}
f_{0}=0.162\frac{t}{L^{2}}v_s \label{f0_cl}
\end{equation}
where $v_s$ is the speed of sound in quartz as defined above.

The calculation of the sensitivity is slightly more complicated
than in the case of the needle sensor. Here, we adapt the result
from \cite{Giessibl2000APL}:
\begin{equation}\label{eq_S_qPlus_k}
S_{qPlus}^{theory}=q_{el}/A=12d_{21}k\frac{L_{e}(L-L_{e}/2)}{t^{2}}.
\end{equation}
Standard qPlus sensors with dimensions listed in table
\ref{table_sensordims} yield $S_{qPlus}^{theory} = 2.8\,\mu $C/m. It is important to note that the calculated
sensitivity assumes a field distribution as shown in Fig. \ref{fig_sensorgeom} e), while
the actual field looks more like Fig. \ref{fig_sensorgeom} c).

\subsection{Sensor based on length extensional resonator (needle sensor)}
The needle sensor consists of two coupled beams that oscillate opposite to each
other (see Figs.\,2a, 3f). The longitudinal stiffness of $k'$ of each of
the two bars that constitute the needle sensor is given by
\begin{equation}\label{LER_k}
    k'=\frac{E w t}{L} ,
\end{equation}
with Young's modulus $E$, length $L$, width $w$ and thickness $t$
of each quartz beam.
The fundamental eigenmode is a longitudinal standing wave with a
node at the root of each beam and its end at a maximal deflection, thus
the length of one beam $L$ corresponds to a quarter wavelength $\lambda/4$.
 Because
the velocity of sound is $v_s=\sqrt{E/\rho}$ with mass density $\rho$, the eigenfrequency
is given by
\begin{equation}\label{LER_f0}
    f_0= \frac{v_s}{4L}.
\end{equation}
The deflection of a cross section at a distance $z$ from the mount
is given by
\begin{equation}\label{LER_deflection}
    \delta z(z)=A\sin(\frac{\pi z}{2L})
\end{equation}
when the ends of the device oscillate at amplitude $A$. The strain
as a function of $z-$position is then given by
\begin{equation}\label{LER eps}
    \epsilon(z)=\frac{\partial \delta z(z)}{\partial
    z}=\frac{\pi A}{2L}\cos(\frac{\pi z}{2L}).
\end{equation}
The strain $\epsilon$ leads to a mechanical stress $\sigma_{mech}$ given by
\begin{equation}\label{LER sigmech}
    \sigma_{mech}(z)=E \epsilon(z).
\end{equation}
The piezoelectric effect causes the
emergence of a surface charge density $\sigma_{el}$ given by
\begin{equation}\label{LER sigel}
    \sigma_{el}(z)=d_{21} \sigma_{mech}(z)
\end{equation}
where $d_{21}=2.31$\,pC/N is the transverse piezoelectric coupling
coefficient of quartz \cite{Ward1992}, which is equal to the longitudinal piezoelectric coupling
coefficient $d_{11}$. It is important to note that $d_{21}$ is essentially constant over the temperature
range from 1.5\,K to room temperature \cite{Ward1992}. When the charge density is integrated over
the surface of the sensor, the total charge $q_{el}$ at a given deflection
$A$ is given by:
\begin{equation}
    q_{el}=d_{21} w \int_{-L_{e}}^{L_{e}} E \frac{A\pi}{2L} \cos(\frac{z\pi}{2L})dz.
\end{equation}
Thus, the sensitivity is given by
\begin{equation}\label{eq_S_needlesens}
    S_{LER}^{theory}=q_{el}/A=2d_{21} E w\sin(\frac{L_e\pi}{2L}).
\end{equation}
With eq. \ref{LER_k}, we can express eq.
\ref{eq_S_needlesens}
\begin{equation}\label{eq_S_needlesens_k}
    S_{LER}^{theory} = 2 d_{21} k' \frac{L}{t}\sin(\frac{\pi L_e}{2L}).
\end{equation}
The electrodes extend almost to the end of the beams ($L_e =
1.1$\,mm, $L = 1.34$\,mm), therefore, the sine in the equation
above is almost one (exact value 0.960685188) and with $L/t =
1340/70$, we find  $S_{LER}^{theory}\approx 19\times d_{21}\times
k' $. With the stiffness $k'=540$\,kN/m, we find a theoretical
sensitivity of $S_{LER}^{theory}=45$\,$\mu$C/m.

\section{Signal}
\subsection{Frequency shift for exponential force laws and amplitude dependence of signal-to-noise ratio}
In FM-AFM, the signal is a frequency shift
$\Delta f$. This frequency shift depends on the tip sample
interaction and the stiffness $k$, eigenfrequency $f_0$ and
amplitude $A$ of the cantilever. For a force that follows an
exponential distance dependence $F(z)=F_0\exp(-\kappa z)$, we find
\begin{equation}\label{eq_df_exponential_Bessel}
\Delta f =\frac{f_0}{kA}F_0e^{-\kappa (z+A)} I_1(\kappa A)
\end{equation}
where $I_1(\kappa A)$ is the Bessel function of the first kind, a
special version of the Kummer function \cite{Giessibl2000}.

As we
will see below, the noise in the frequency measurement of the
sensor is inversely proportional to $A$, therefore the
signal-to-noise ratio (\textit{SNR}) can be expressed as
\begin{equation}
\textit{SNR} \propto e^{-\kappa A} I_1(\kappa A).\label{SNR}
\end{equation}

\begin{figure}
\begin{center}
\includegraphics[clip=true, width=0.8\textwidth]{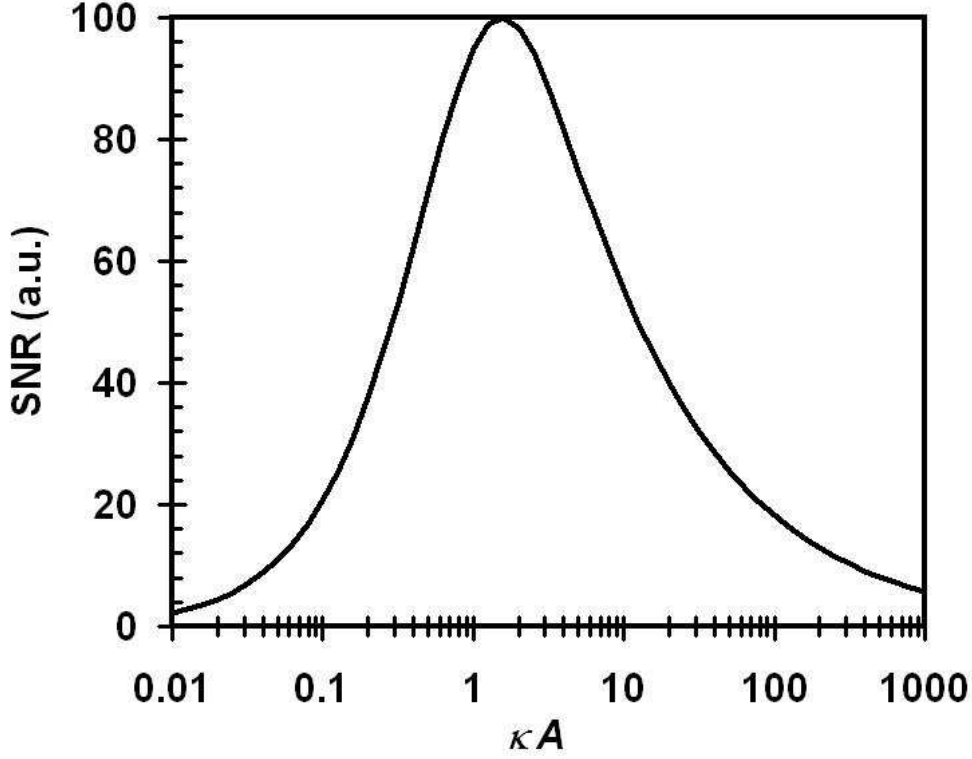}
\end{center}
\caption{Signal-to-noise ratio (SNR) as a function of the product between decay
constant $\kappa$ and amplitude $A$, where the decay constant $\kappa$ is inverse to
the interaction length $\lambda$, thus $\kappa=1/\lambda$. Optimal SNR is obtained for $\kappa A = A/\lambda = 1.545.$
}\label{fig_SNR}
\end{figure}
This function has its maximum at $\kappa A=1.5451...$, thus, the
optimal signal-to-noise ratio is reached for amplitudes that
correspond to the decay length $\lambda=1/\kappa$ of the
tip-sample force \cite{Giessibl1999ASS}, or more precisely
$A_{opt} \approx 1.545 \lambda$. In theory, this ideal amplitude
applies to all sensors in FM-AFM that probe interactions of range
$\lambda$, provided the sensor stiffness is sufficient to enable
stable oscillation close to the surface \cite{Giessibl1999ASS}.

We can rewrite eq. \ref{eq_df_exponential_Bessel} such that its
resemblence to the gradient approximation becomes more clear:
\begin{equation} \label{eq_dfBesselgrad}
\Delta f =\frac{f_0}{2k} \kappa F_0 e^{-\kappa z}
\frac{2I_1(\kappa A)e^{-\kappa A}}{\kappa A}.
\end{equation}
The first factor in this equation is the gradient approximation,
while the fraction $2I_1(x)e^{-x}/x$ with $x=\kappa A$ can be
expanded as $2I_1(x)e^{-x}/x = 1-x+5/8 x^2+O(x^3)$.
 For a minimum distance between tip and sample of $z$, the
tip oscillates within the interval $[z..z+2A]$ and at the optimal
oscillation amplitude $A_{opt} \approx 1.545/ \kappa$, we obtain
an average tip-sample force gradient that is approximately one
third of the peak force gradient at distance $z$, because $
2I_1(1.5451)e^{-1.5451}/1.5451\approx 0.33$.
\subsection{Frequency shift for a tip-sample force modelled by a Morse potential}
\begin{figure}
\begin{center}
\includegraphics[clip=true, width=0.8\textwidth]{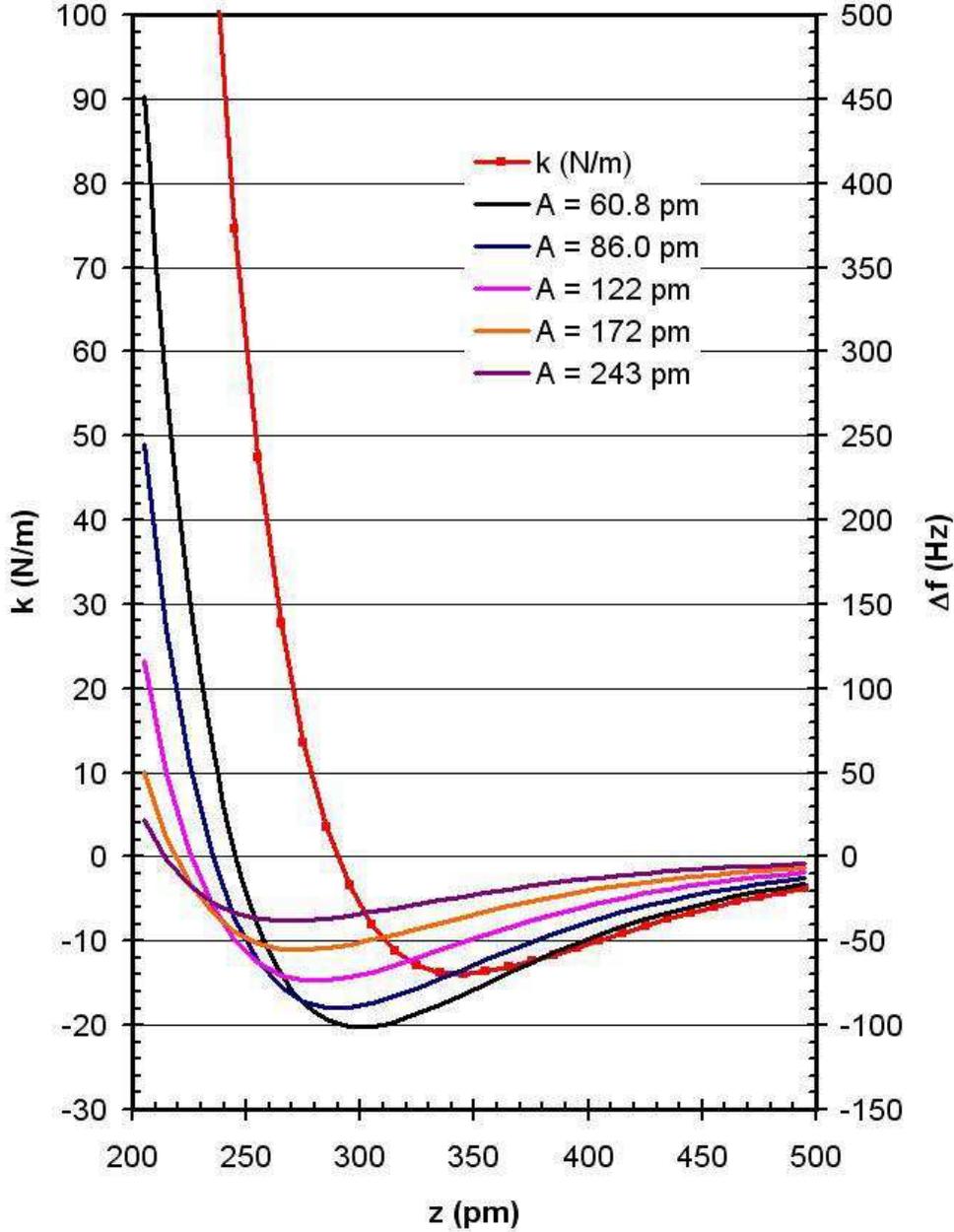}
\end{center}
\caption{(Color online) Force gradient (red) and calculated
frequency shift for the interaction of a silicon tip with an
adatom on Si(111)-(7$\times$7) surface modelled by a Morse
potential with $E_{bond}=2.273$\,eV, $\sigma=235.7$\,pm and
$\kappa=12.70$\,nm$^{-1}$ for a qPlus sensor with $k=1800$\,N/m
and $f_0=30$\,kHz and various amplitudes (see legend). If a
standard needle sensor was used here, the frequency shift values
denoted on the right vertical axis have to be multiplied by 1/20,
because the frequency shift is proportional to $f_{0}/k$. For the
qPlus sensor, a minimal frequency shift of -70\,Hz results at the
optimal amplitude $A=122$\,pm, while the needle sensor only yields
a minimal frequency shift of -3.5\,Hz. }\label{fig_deltafMorse}
\end{figure}

We can now calculate the frequency shift assuming that a single
chemical bond is responsible for the contrast. A covalent bond
between a Si tip atom and an adatom on Si(111)-(7$\times$7) can be
modelled by a Morse Potential
\begin{equation}
V_{Morse}=E_{bond}(-2e^{-\kappa(z-\sigma)}+e^{-2\kappa(z-\sigma)})
\label{Morse}
\end{equation}
with the following fitting parameters: bond strength
$E_{bond}=2.273$\,eV, equilibrium distance $\sigma=235.7$\,pm and
decay constant $\kappa=2\times
1.497/0.2357$\,nm$^{-1}=12.70$\,nm$^{-1}$ \cite{Perez1998}. The
optimal  amplitude to measure this bond in the attractive regime
is therefore $A_{opt} = 1.545/12.7$\,nm = 122\,pm. The repulsive
regime of this bond would ideally be probed with an amplitude of
61\,pm, because the range of the repulsive force component is only
half the range of the attractive component. Figure
\ref{fig_deltafMorse} displays the force gradient and the
frequency shifts corresponding to a sensor that oscillates in a
force field given by this Morse potential.

Figure \ref{fig_deltafMorse} shows that at the optimal oscillation
amplitude, a minimal frequency shift of -70\,Hz can be expected
for a standard qPlus sensor and -3.5\,Hz for a standard needle
sensor when probing a single silicon bond. However, on weekly
bonding systems such as organic molecules, absolute frequency shifts on the
order of -5\,Hz \cite{GrossScience2009b} for a qPlus sensor with a contrast on the order of
0.1\,Hz result. A needle sensor would change its
frequency by only -250\,mHz with a contrast of about 3\,mHz for the same interaction.

\section{Noise}\label{sec_noise}

If the frequency of the force sensor could be measured with
infinite accuracy, infinitely small force gradients could be
measured. In practice, there are four relevant noise contributions
that need to be considered. For large bandwidths, i.e. for high
scanning speeds, deflection detector noise is dominant. Deflection detector noise increases
with $B^{3/2}$. Two other noise sources, thermal noise and
oscillator noise, increase with the square root of bandwidth $B$.
The forth noise source is due to sensor frequency drifts caused by
temperature changes. Thermal frequency drift is a challenge for room temperature
measurements and in particular for high-temperature measurements. Because
we measure an average force gradient in FM-AFM, the noise in this figure is
given with eq. \ref{gradient_approximation}
\begin{equation}
\delta k_{ts} = 2k  \frac{\delta f}{f_{0}}.\label{eq_noise_kts_vs_df}
\end{equation}

\subsection{Deflection detector noise}
The deflection of the cantilever can not be measured with infinite
precision, but is subject to noise. Typically, the oscillation frequency of the
cantilever varies very little around the eigenfrequency $f_0$ and we can
therefore assume a constant deflection detector noise density $n_q$ that denotes the precision at which the deflection
of the cantilever can be measured (e.g. for $n_q=100$\,fm$/\sqrt{\textrm{Hz}}$, the error in deflection measurement is
$\delta q = 100$\,fm at a bandwidth of 1\,Hz and $\delta q = 1$\,pm at a bandwidth of 100\,Hz). This
uncertainty in the deflection measurement also leads to frequency
noise \cite{Eguchi2002,Hasegawa2004,Kobayashi2009}, given by
\begin{equation}
\frac{\delta f_{det}}{f_{0}}= \sqrt{\frac{2}{3}} \,\frac{
n_{q}B^{3/2}}{Af_0}. \label{df_sensor}
\end{equation}
With eq. \ref{eq_noise_kts_vs_df}, we find
\begin{equation}
\delta k_{ts\,det} = \sqrt{\frac{8}{3}} \frac{k n_{q}}{f_0}
\frac{B^{3/2}}{A}. \label{dkts_sensor}
\end{equation}
For quartz sensors, the deflection noise depends on the charge
that is generated per deflection and the gain and noise of the
preamplifier. Current-to-voltage converters convert the current
provided by the quartz sensor to a voltage. However, the frequency
response of the current-to-voltage converter is not independent of frequency, but given by
\begin{figure}
\begin{center}
\includegraphics[clip=true, width=0.8\textwidth]{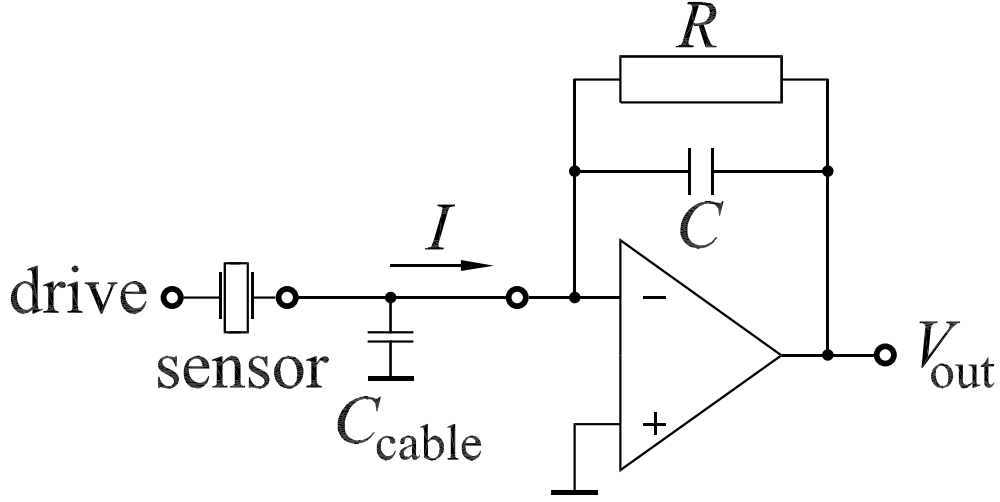}
\end{center}
\caption{Schematic of a quartz sensor, cable and current-to-voltage converter that is often
used for amplifying deflection data from quartz sensors. The gain of the amplifier
is given by $V_{out}=-R I/(1+i f/f_{c1})$ with its first corner frequency $f_{c1}$ given by $f_{c1}=1/(2\pi RC)$.
The capacity of the cable should be as low as possible - cable capacity increases
noise in the amplifiers output. If the amplifier is vacuum compatible, it can be placed
close to the sensor, thus reducing cable capacity and noise. The sensor can be excited electrically as shown
in this figure or mechanically - the drive signal is grounded in this case.} \label{fig_amp}
\end{figure}
\begin{equation}
V_{out}= -\frac{R I}{1+i 2\pi f RC}, \label{eq_gain_f_ivc}
\end{equation}
where $R$ is the resistance of the feedback resistor and $C$ is its parasitic capacitance.
The red line in figure \ref{fig_gain_vs_f} shows the theoretical
frequency response of an ideal operational amplifier with $R=100$\,M$\Omega$ and a
parasitic capacitance of $C=0.2$\,pF. The gain is flat for frequencies smaller than
the corner frequency $f_{c1}=1/(2\pi RC)=7.96$\,kHz.
For $f>>f_{c1}$, the gain is given by $V_{out}= -I/(i 2\pi f C)$ - inversely proportional to $f$.
A sinusoidally varying charge $Q_{ch}=Q_0\exp{(i2\pi f t)}$ corresponds to a current
$I=\dot{Q}_{ch} =Q_0 i 2\pi f \exp{(i2\pi f t)}$, thus the gain can be
expressed as $V_{out}= -Q_{ch}/C$. Therefore, this amplifier is called a
\lq charge amplifier\rq{} for frequencies significantly larger than $f_{c1}$. Simple
amplifiers as the one shown in Fig. \ref{fig_amp} often display a second corner frequency $f_{c2}$
not very much higher than $f_{c1}$ and for frequencies beyond $f_{c2}$
the gain decays proportional to $1/f^2$. The charge amplifier that is used here
for the needle sensor (Kolibri-amplifier \cite{Kobayashi2009}, \cite{Femto}) has an $f_{c2}$ at around 15\,MHz and is therefore suited well for
high-frequency sensors.
\begin{figure}
\begin{center}
\includegraphics[clip=true, width=0.8\textwidth]{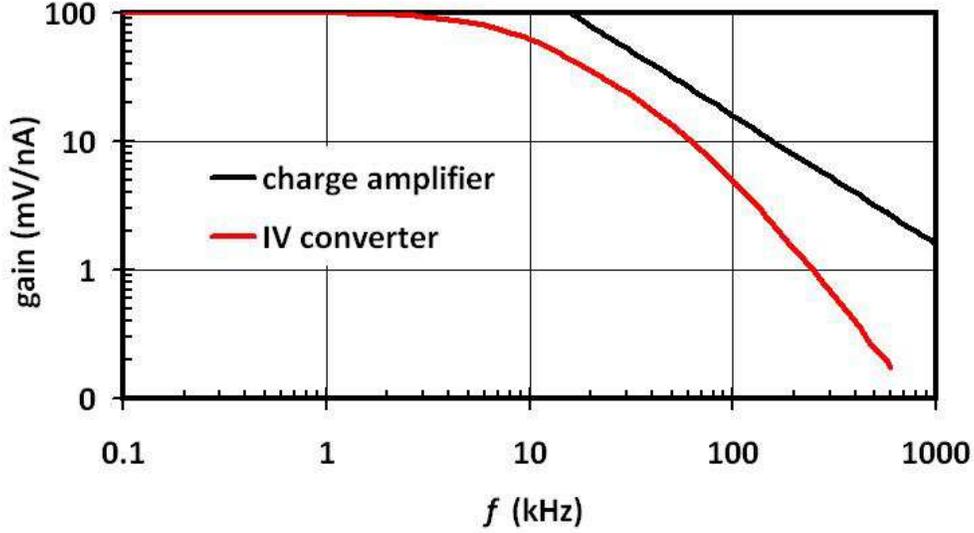}
\end{center}
\caption{(Color online) Current gain versus frequency for a
current-to-voltage converter built from an ideal operational
amplifier and a 100\,M$\Omega$ feedback resistor with a parasitic
capacitance of 0.2\,pF (red line), yielding a first corner
frequency (here, $f_{c1}=8$\,kHz). For frequencies higher than
$f_{c1}$, the gain drops proportional to $1/f$. Typically, these
simple amplifiers develop a second corner frequency (here
$f_{c2}=80$\,kHz) \cite{deFariaElsner2010}, for frequencies higher
than $f_{c2}$, their gain drops proportional to $1/f^2$. The
black line displays the gain of a commercial charge amplifier
\cite{Femto} with a constant gain of $10^{13}$\,V/C (black line)
for a remarkably large frequency range from 250\,Hz to 15\,MHz.}
\label{fig_gain_vs_f}
\end{figure}
The question is now, when is it advisable to use a current-to-voltage converter, and when is it favorable to use
a charge amplifier. Figure \ref{fig_gain_vs_f} shows that the current-to-voltage converter becomes a charge amplifier
for sufficiently large frequencies. While one can increase $f_{c1}$ by reducing the value of the feedback resistor $R$, a reduction
of $R$ increases the current noise. The tradeoff between noise and bandwidth leads to an optimal amplifier type for a given operating
frequency. Here, we found out that our home-built current-to-voltage converter has a better signal-to-noise ratio for frequencies around
$(30\pm 10)$\,kHz, while the FEMTO amplifier \cite{Femto} works better for frequencies above.
For charge amplifiers, the deflection detector noise density can be expressed by
\begin{equation}
n_q= \frac{n_{amp}}{S} \label{eq_nqnamp}
\end{equation}
where $n_{amp}$ is the noise density of the preamplifier and $S$
is the sensitivity (charge per deflection) as calculated for the
needle sensor in eq. \ref{eq_S_needlesens_k} and for the qPlus
sensor in eq. \ref{eq_S_qPlus_k}.
\begin{equation}
\delta k_{ts\,det} = \sqrt{\frac{8}{3}} \frac{k }{S f_0}n_{amp}
\frac{B^{3/2}}{A}. \label{dkts_sensor_S}
\end{equation}
This equation shows, that the deflection detector noise is small
for small spring constants, small amplifier noise, large sensitivity and
large eigenfrequency. Thus, the figure of merit for the sensor is not $S$ alone, but $S f_0/k$.
For both needle and qPlus sensors, the sensitivity is proportional to $k$. We find for the needle sensor
\begin{equation}
\delta k_{ts\,det\,ns}= \sqrt{\frac{8}{3}}  \frac{ n_{amp}t
B^{3/2}}{d_{21}LAf_0} \label{dkts_needlesensor}
\end{equation}
for the ideal case of $L_e=L$. For the qPlus sensor, we find
\begin{equation}
\delta k_{ts\,det\,qP}= \sqrt{\frac{8}{3}}  \frac{
n_{amp}t^2 B^{3/2}}{6d_{21}L^2Af_0}, \label{dkts_qPlus}
\end{equation}
again assuming the ideal case of $L_e=L$.
Thus, deflection detector noise depends on the properties of the sensor
and the amplifier. If we assume a charge noise density of
$n_{amp}=90$\,zC/$\sqrt{\text{Hz}}$ (such as achieved by the commercial
FEMTO amplifier \cite{Femto} when loaded with a 1\,m
coaxial cable corresponding to a 100\,pF cable capacitance), we can now calculate
an explicit number for the deflection detector noise contribution to the force
gradient noise with $A=100$\,pm and the geometrical values after
table \ref{table_sensordims}. For the needle sensor, we find a theoretical deflection detector noise contribution of
\begin{equation}
\delta k_{ts\,needle\,sensor}= 33.2 \mu\textrm{N/m}\frac{
B^{3/2}}{\textrm{Hz}^{3/2}}
\end{equation}
and for the qPlus sensor, we find a theoretical deflection detector noise contribution of
\begin{equation}
\delta k_{ts\,qPlus\,sensor}= 25.7 \mu\textrm{N/m}\frac{
B^{3/2}}{\textrm{Hz}^{3/2}}
\end{equation}
For a bandwidth of 100\,Hz, the theoretical deflection detector noise contribution is thus
33.2\,mN/m for the needle sensor and 25.7\,mN/m for the qPlus
sensor. However, we have based this calculation on the theoretical
sensitivity of the sensors, we will see further below that while the experimental sensitivity of the needle sensor
matches theory, the qPlus sensor develops only about 50\,\% of the theoretical sensitivity.
Deflection detector noise depends dramatically on bandwidth, it can be
reduced substantially by bandwidth reduction. At low temperatures,
where slow scanning is possible, the bandwidth can be reduced to
one Hertz or less and tiny force gradients can be detected in this
case. For a bandwidth of 1\,Hz, the deflection detector noise contribution is thus
33.2\,$\mu$N/m for the needle sensor and 25.7\,$\mu$N/m for the qPlus
sensor. However, at low bandwidth the remaining three noise sources are typically much larger
than deflection noise.

\subsection{Thermal noise}
The thermal noise of a force sensor at a bandwidth $B$ is given by \cite{Albrecht1991}:
\begin{equation}
\frac{\delta f_{thermal}}{f_{0}}= \sqrt{\frac{k_{B}TB}{\pi k A^{2}
f_{0} Q}}.  \label{df_thermal}
\end{equation}
Thus, the thermal noise in force gradient measurement is given by
\begin{equation}
\delta k_{ts\,thermal}=\sqrt{\frac{4 k k_{B}TB}{\pi A^{2} f_{0} Q}}
\propto \sqrt{\frac{ k }{ f_{0} Q}}. \label{delta_kts_thermal}
\end{equation}
For the needle sensor, reasonable $Q$ values are 15\,000 at room temperature and
80\,000 at 4\,K \cite{An2008}. For the qPlus sensor, $Q\approx 3\,000$ at room temperature, reaching up to
200\,000 at 4\,K \cite{Hofmann2011}. Thus, at room temperature the thermal contribution to
the minimal detectable force gradient is $\delta k_{ts\,thermal}=6$\,mN/m per $\sqrt{\textrm{Hz}}$ for the
needle sensor and $\delta k_{ts\,thermal}=3$\,mN/m per $\sqrt{\textrm{Hz}}$
for the qPlus sensor. At $T=4$\,K, the minimal detectable force gradient is $\delta k_{ts\,thermal}=390$\,$\mu$N/m per $\sqrt{\textrm{Hz}}$ for the
needle sensor and $\delta k_{ts\,thermal}=40$\,$\mu$N/m per $\sqrt{\textrm{Hz}}$
for the qPlus sensor. Again, these calculations refer to $A=100$\,pm.

\subsection{Oscillator noise}

Recently, Kobayashi et al. \cite{Kobayashi2009} discovered a new
contribution to frequency noise in FM-AFM that arises in
particular in low $Q$ environments. However, this contribution is not
explicitly temperature dependent and thus can
become significant at low temperatures where thermal noise becomes small.
The
origin of this noise can be understood as a driving of the
cantilever off resonance because the amplitude feedback is fed
with a noisy input signal (due to a finite $n_q$). The lower the
$Q$- value, the more of this noise pushes the cantilever at the
correct phase, therefore, this noise contribution is proportional
to $n_q$ and inversely proportional to $Q$:
\begin{equation}
\frac{\delta f_{osc}}{f_{0}} = \frac{n_q B^{1/2}}{\sqrt{2} A Q}.
\end{equation}
With eq. \ref{eq_noise_kts_vs_df}, we find
\begin{equation}
\delta k_{ts\,osc} = \sqrt{2} \frac{k n_q }{Q}\frac{B^{1/2}}{ A }.
\label{delta_kts_osc}
\end{equation}
Similar to thermal noise, oscillator noise is proportional to the
square root of the detection bandwidth $B$ and inversely
proportional to amplitude. For the $Q$ values from above, we find room temperature
values of
$\delta k_{ts\,osc}=4.6$\,mN/m per $\sqrt{\textrm{Hz}}$ for the
needle sensor and $\delta k_{ts\,thermal}=0.6$\,mN/m per $\sqrt{\textrm{Hz}}$
for the qPlus sensor. At $T=4$\,K, the contribution of oscillator noise to the minimal detectable force gradient is $\delta k_{ts\,osc}=1.4$\,mN/m per $\sqrt{\textrm{Hz}}$ for the
needle sensor and $\delta k_{ts\,thermal}=9.5$\,$\mu$N/m per $\sqrt{\textrm{Hz}}$
for the qPlus sensor. Again, these calculations refer to $A=100$\,pm.

\subsection{Thermal frequency drift noise}

\begin{figure}
\begin{center}
\includegraphics[clip=true, width=0.5\textwidth]{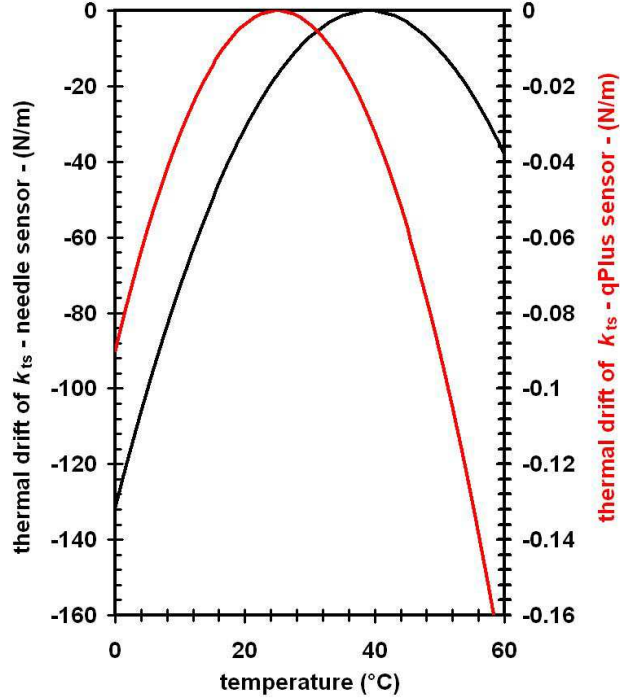}
\end{center}
\caption{(Color online) Effect of temperature changes on the
measured tip-sample force gradient. Both needle sensor and qPlus
sensor change their frequency as a function of temperature.
Although the relative frequency shift is much smaller than for
silicon cantilevers, the effect on the measured force gradient
scales with stiffness $k$. This thermal frequency drift noise is
almost three orders of magnitude smaller for the qPlus sensor than
for the needle sensor.}
\end{figure}
Temperature variations cause a drift in eigenfrequency. For
silicon cantilevers, the relative frequency variation is linear
with temperature with a value of $-35$\,ppm/K at room temperature
\cite{Gysin2004}. Thus, a hypothetical Si cantilever with $k=1$\,kN/m (this large
stiffness would be required to enable stable oscillation at small amplitudes)
would be subject to a $\langle k_{ts}\rangle$ drift of $-35$\,mN/m/K.
Quartz sensors show a quadratic frequency shift
with temperature and the eigenfrequency varies with temperature
as an inverted parabola centered around the \textit{turnover
temperature} $T_{p}$ \cite{MicroCrystal}:
\begin{equation}\label{eq_temp_drift_f}
\frac{\delta f_{sensor}}{f_{0}} = -\chi  (T-T_{p})^2.
\end{equation}
The turnover frequency depends on the crystal cut (see Fig. 9 in \cite{Momosaki1985}). Tuning fork
crystals are often cut to yield $T_p=298$\,K such that the turnover temperature is close
to the temperature that a watch that is strapped to a wrist typically develops. Length-extensional-resonators, in contrast,
are often oriented such that their turnover temperature is around 313\,K \cite{MicroCrystal}, probably because 1\,MHz crystals
are typically not worn on the wrist but built into printed circuit boards that have higher operating temperatures than the human body. Here we chose an LER with $T_p=298$\,K to be able to compare the frequency drift of both types of sensors at room temperature.
This thermal frequency drift causes a thermal drift in force
gradient measurement given by
\begin{equation}\label{eq_temp_drift_kts}
\delta k_{ts\,drift} = -2k\chi (T-T_{p})^2.
\end{equation}
Although the temperature stability of quartz is excellent with
very small values of $\chi=35 \times 10^{-9}$K$^{-2}$ \cite{MicroCrystal}, the net
effect on the precision on the measurement of $\langle
k_{ts}\rangle$ is proportional to the effective stiffness of the
sensor $k$.

The quadratic dependence of the frequency variation with
temperature is only valid for temperatures around $T_{p}$. For
the temperature range from 300\,K to 4\,K, the frequency variation has been measured by
Hembacher et al. \cite{Hembacher2002ASS} and is approximately given by

\begin{equation}\label{eq_temp_drift_f_approx}
\frac{\delta f_{sensor}}{f_{0}} \approx -0.00081 [1-\cos
((T/T_{p}-1)\pi)]
\end{equation}
with a total relative frequency change of $-1620$\,ppm over
the temperature range from 300\,K to 4\,K. An et al. have
found a similar frequency change of a needle sensor (Fig. 3 in \cite{An2008})
from 998066\,Hz at 300\,K to 996314\,Hz, corresponding to $-1755$\,ppm.
This equation shows that frequency drift with temperature is
particularly large for temperatures between room temperature and
absolute zero. This approximate formula models the
data measured by Hembacher et al. \cite{Hembacher2002ASS} quite
precise down to liquid helium temperatures.
Because the
relative frequency shift is mainly dependent on the variation of
the velocity of sound with temperature (pp. 38 in \cite{Morita2002}), we
expect a similar relative frequency shift for the qPlus sensor and
the needle sensor also in the whole temperature range from 0\,K to 300\,K.
\begin{figure}
\begin{center}
\includegraphics[clip=true, width=0.5\textwidth]{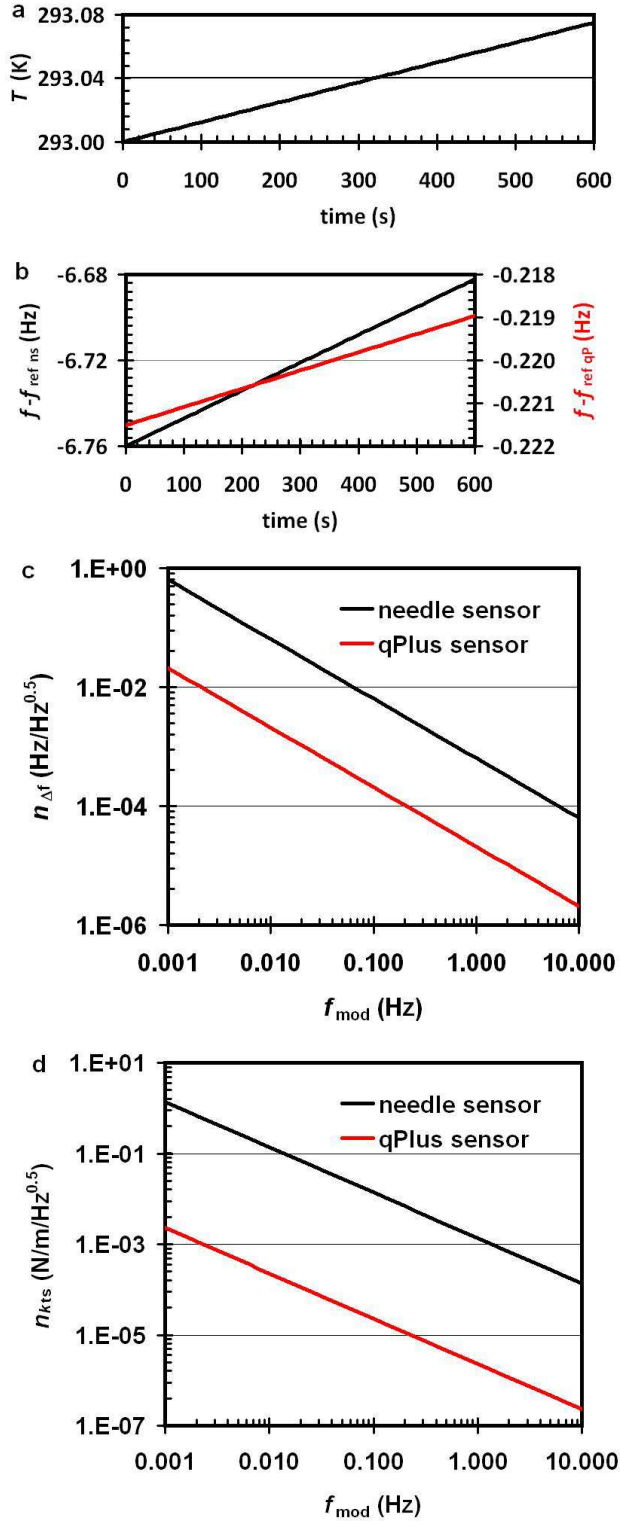}
\end{center}
\caption{(Color online) Effect of temperature drift on frequency drift, frequency noise at the PLL output and
force gradient noise. a) A temperature drift of 125\,$\mu$K/s is assumed, yielding a temperature increase of 75\,mK over ten minutes.
b) Frequency drift at at temperature 10\,K above or below the turnover temperature $T_p$, see eq. \ref{eq_temp_drift_f}. For the needle sensor,
the absolute frequency change over 10 min is 78\,mHz, while for the qPlus sensor, it is 2.5\,mHz. c) Power spectral density of the frequency drift noise for
needle and qPlus sensor. A linear frequency drift with time causes a $1/f$ power spectrum. d) Power spectral density of the tip-sample force gradient
noise due to drift. This noise contribution is linear with the force constant of the sensor, i.e. it is 600 times larger for the needle sensor than for the
qPlus sensor.
}\label{fig_fdriftnoise}
\end{figure}
We now analyze the effect of temperature drift on the measured tip sample force gradient.
First, we look at the frequency drift of the sensor for a given rate of temperature change.
Figure \ref{fig_fdriftnoise} a) shows temperature versus time for a constant drift rate of
$dT/dt=125\mu$K/s at $T-T_p=10$K over a time interval of 10 minutes. The frequencies of quartz sensors vary according to eq. \ref{eq_temp_drift_f}
by a rate $r_{ns}= 100\,\mu$Hz/s for the needle sensor and $r_{qP}= 3.3\,\mu$Hz/s for the qPlus sensor.

Now, we can compute the power spectral density of the frequency drift noise contribution by taking a Fourier transform of the
square of the frequency drift. The reason we are not just adding the frequency noise contributions but adding the squares is that detector, thermal, oscillator and thermal drift noise
are statistically independent and the net effect of statistically independent variables is computed by taking the square root of the sum of squares.
For a frequency drift that is linear with time, we find $\delta f(t)=r\times t$ within a time interval $[-\tau/2..\tau/2]$.
With $\Omega=2\pi/\tau$, we can express the time dependence of the frequency as
\begin{equation}\label{eq_fourier_series_f}
\delta f^2(t)=\sum_{n=0}^{\infty}a_n \cos{(n\Omega t)}
\end{equation}
with Fourier coefficients
\begin{equation}
a_n=\frac{\Omega}{\pi}\int_{t=-\tau/2}^{\tau /2} r^2t^2\cos{(n\Omega t)}dt
\end{equation}
and
\begin{equation}\label{eq_fourier_a_n}
a_n=(-1)^n\frac{r^2\tau^2}{\pi^2n^2}.
\end{equation}
We can now interpret $|a_n|$ as the equivalent power component at a frequency $f_{mod}=n/\tau$ in a frequency interval of $1/\tau$.
Therefore, the power spectral density (power per frequency) becomes
\begin{equation}
n_{\Delta f\,drift}^2(f_{mod})=\frac{r^2\tau}{\pi^2f_{mod}^2}
\end{equation}
and
\begin{equation}\label{eq_n_fmod_drift}
n_{\Delta f\,drift}(f_{mod})=\frac{r\sqrt{\tau}}{\pi f_{mod}}.
\end{equation}
Thus, a linear frequency drift leads to $1/f$ noise in the frequency spectrum of the PLL output. The magnitude of this
noise component depends on the drift rate of the frequency $r$ and the measurement period $\tau$. The time period $\tau$ is at least the time
it takes to complete one image. Thus, for fast measurements, frequency drift
noise can be reduced provided that the frequency detector (PLL) is reset before an image is taken.
To obtain the effect of this noise on the force gradient measurement, we need to multiply $n_{\Delta f}(f_{mod})$
by $2k/f_0$ (see eq. \ref{gradient_approximation}) to obtain
\begin{equation}\label{eq_n_kts_drift}
n_{kts\,drift}(f_{mod})=\frac{2 k r\sqrt{\tau}}{f_0 \pi f_{mod}}.
\end{equation}
Because the frequency drift rate is proportional to $f_0$, the force gradient noise due to
thermal drift is proportional to the stiffness of the sensor $k$, and thus this noise source is
600 times larger for the needle sensor than for the qPlus sensor.

\subsection{Summary of noise calculations}
In summary, we find that the large spring constant of the needle sensor is not a significant disadvantage regarding deflection deflection detector noise,
because although the frequency
shift that a sensor is subject to is proportional to $1/k$, the sensitivity is proportional to $k$, and
the two effects cancel. However, $k$ does affect the other three noise sources: thermal
noise increases as $\sqrt{k}$, and both oscillator noise and frequency drift noise are proportional to $k$.
Therefore, the recommendations in eq. \ref{eq_k_opt}, stating that $k$ should be large enough to enable stable
sensor oscillations at the optimal amplitude but otherwise as small as possible are still valid.
High $Q$-values are desirable to minimize thermal and oscillator noise. The frequency drift noise can be minimized by
operating the sensors in a thermally stable environment, preferentially at temperatures at or close to $T_{p}$.

\section{Experimental noise measurements}
\subsection{Deflection spectrum at thermal excitation}
\begin{figure}
\begin{center}
\includegraphics[clip=true, width=0.8\textwidth]{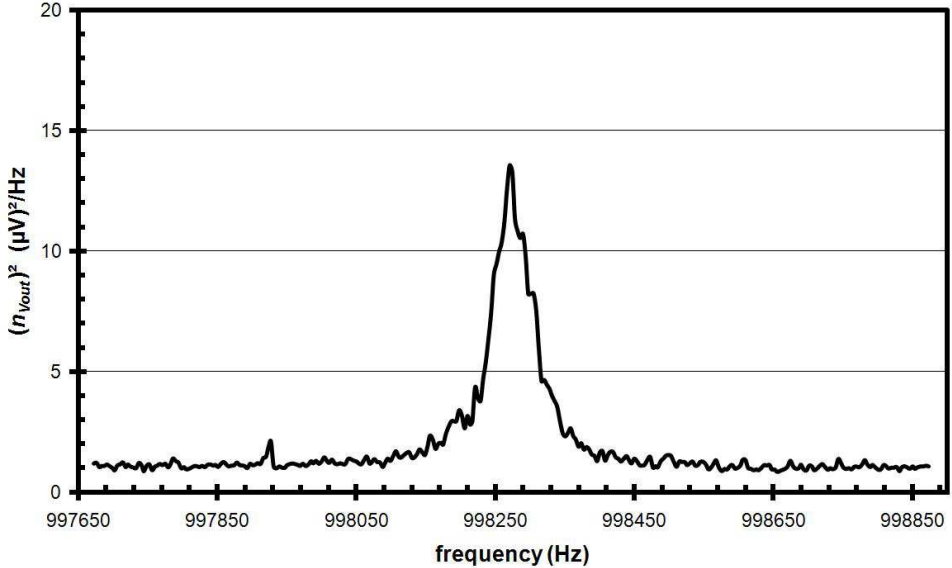}
\end{center}
\caption{Thermal spectrum of a needle sensor with standard
dimensions at room temperature and ambient pressure. A commercial preamplifier \cite{Femto} was used. The sensitivity
of the sensor is calculated to 45.4\,$\mu$C/m, the $Q$-factor is 18500 and the deflection detector noise density is 1.89\,fm/$\sqrt{\textrm{Hz}}$.}\label{fig_nvoutns}
\end{figure}
So far, we have only considered theoretical calculations to
compare the noise characteristics of the two sensors studied here.
Now, we supplement the calculations by measurements. First, we
measure the thermal noise peak of the needle sensor and the qPlus
sensor with sensors of standard dimensions listed in table
\ref{table_sensordims}. The equipartion theorem states, that an
oscillator carries a thermal energy $k_BT/2$ per degree of
freedom, where $k_B$ is Boltzmann's constant and $T$ is the
temperature in Kelvin. For the standard qPlus sensor, we find the
thermal amplitude by equating the average potential energy to the
thermal energy $kA_{rms}^2/2=k_BT/2$, yielding a thermal
rms-amplitude of $A_{rms}=1.52$\,pm or peak-amplitude of
$A_{0\,p}=2.14$\,pm. For the needle sensor, we need to take into
account that it is a coupled oscillator, therefore $2\times
k'A_{rms}^2/2=k_BT/2$, yielding a thermal rms-amplitude of
$A_{rms}=62$\,fm or peak-amplitude of $A_{0\,p}=88$\,fm.
Figure \ref{fig_nvoutns} shows the thermal peak of a needle sensor without tip
in ambient conditions. The power spectral density in Fig. \ref{fig_nvoutns} was recorded by connecting
the output of the FEMTO amplifier \cite{Femto} to the input of the oscillation controller (OC4 from Nanonis \cite{Nanonis}) using
the Zoom-FFT (Fast Fourier Transform) feature and correcting the filter error by comparing the output
with a dedicated FFT Analyzer at low frequencies (Agilent 35670A Dynamical Analyzer). The input of the FEMTO amplifier was connected to
a length extensional resonator (no tip attached) with dimensions given by table \ref{table_sensordims} with a coaxial cable with a length
of 1\,m (capacity approx. 100\,pF). The
commercial preamplifier has a noise density of
$n_{amp}=90$\,zC/$\sqrt{\text{Hz}}$ when loaded with a 1\,m
coaxial cable (100\,pF cable capacitance) \cite{Femto} and
$n_{amp}=40$\,zC/$\sqrt{\text{Hz}}$ without cable (sensor directly
connected to the amplifier) at the operating frequency of the
needle sensor (1\,MHz). From figure \ref{fig_nvoutns}, we can
calculate the sensitivity as well as the deflection detector noise density
by following the procedure published in \cite{Giessibl2000APL}.

For the needle sensor, we find an experimental sensitivity of $S_{needle\,sensor}^{exp}=45.4 \mu$C/m, that is 100\%
of the theoretical value. In a previous measurement, the needle sensor reached only
44\% of the theoretical value \cite{An2005}. A
possible reason for a deviation between theoretical and
experimental sensitivity in the previous measurement might be attributed to cable capacity
between sensor and amplifier and non-ideal amplifier performance.
The deflection detector noise density is thus
$n_q = 2$\,fm/$\sqrt{\text{Hz}}$ with a 1\,m cable and $n_q = 0.89$\,fm/$\sqrt{\text{Hz}}$ when the sensor
is directly connected to the preamp (not feasible for vacuum operation).

At 30\,kHz, the operating frequency of the
qPlus sensor, we measured $n_{amp}=122$\,zC/$\sqrt{\text{Hz}}$
with a 1\,m coaxial cable (100\,pF cable capacitance) for the FEMTO amplifier \cite{Femto}
and $n_{amp}=86$\,zC/$\sqrt{\text{Hz}}$ without cable. Thus, a
standard qPlus sensor would yield $n_q =
$122\,zC/$\sqrt{\text{Hz}}$/$1.44\,\mu $C/m = 85
fm/$\sqrt{\text{Hz}}$. When directly
connected  to the commercial amplifier, the qPlus sensor would achieve a
deflection detector noise density of $n_q = 60$\,fm/$\sqrt{\text{Hz}}$ at room temperature.
Using our home-built amplifier for the
qPlus sensor, we obtain a deflection detector noise density of $n_q =
62$\,fm/$\sqrt{\text{Hz}}$. The homebuilt amplifier is a current-to-voltage converter based on
a OPA 657 operational amplifier with a feedback resistance of 100\,$\textnormal{M}\Omega$ \cite{Morawski2011}. It is
UHV compatible and therefore can be connected closely to the
sensor, thereby greatly reducing $C_{cable}$. At low temperatures the homebuilt amplifier
can be cooled, and its noise at 4\,K typically drops to 50\% \cite{Hembacher2002ASS},
yielding $n_q = 31$\,fm/$\sqrt{\text{Hz}}$ at 4\,K.
\begin{figure}
\begin{center}
\includegraphics[clip=true, width=0.8\textwidth]{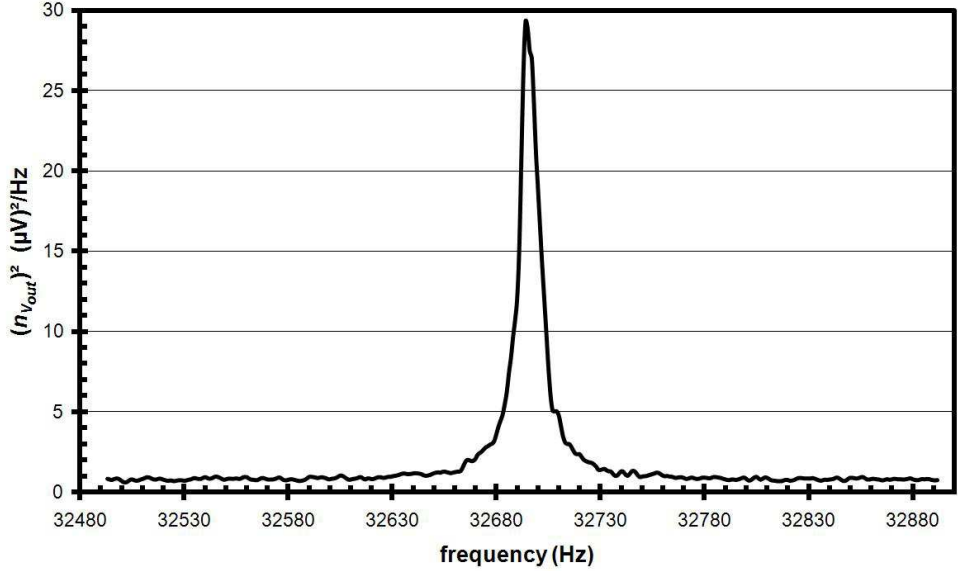}
\end{center}
\caption{Thermal spectrum of a qPlus sensor with standard
dimensions at room temperature and ambient pressure. A homebuilt preamplifier was used. The sensitivity
of the sensor is calculated to 1.44\,$\mu$C/m, the $Q$-factor is 2900 and the deflection detector noise density is 62\,fm/$\sqrt{\textrm{Hz}}$.}
\label{fig_n_vout_qp}
\end{figure}

\subsection{Power spectral density of the frequency detector output}
When the sensor is operating in the AFM, it is excited at a
constant amplitude, and the frequency of the sensor is measured as
the physical observable that relates to the tip-sample forces.
 The power spectral density in Fig. \ref{fig_nvoutns} was
recorded by connecting the output of a home-built UHV compatible
amplifier to the input of the Nanonis OC4 PLL and recording its
FFT (Fast Fourier Transform) output at sufficiently fast settings (demodulation bandwith 1300 Hz,
lock range 305 Hz). The input of the amplifier
was connected to a qPlus sensor without tip with dimensions given
by table \ref{table_sensordims} with a short cable with a length
of approx. 0.1\,m (capacity approx. 10\,pF). The experimental
result is $S_{qPlus}^{exp}  = 1.44\,\mu $C/m - about 51\,\% of the
theoretical value. The deviation between the theoretical and
experimental values is probably due to edge effects - the
calculation of the sensitivity is based on a homogenous field
distribution and an electrode configuration in the quartz crystal
as in Fig. \ref{fig_sensorgeom} e, while the actual field
distribution is perturbed by edge effects as in Fig.
\ref{fig_sensorgeom} c. For the needle sensor, the deviation
between actual (Fig. \ref{fig_sensorgeom} h) and ideal field (Fig.
\ref{fig_sensorgeom} j) is much smaller, therefore its
experimental sensitivity is essentially equal to the calculated
sensitivity.

\begin{figure}
\begin{center}
\includegraphics[clip=true, width=0.8\textwidth]{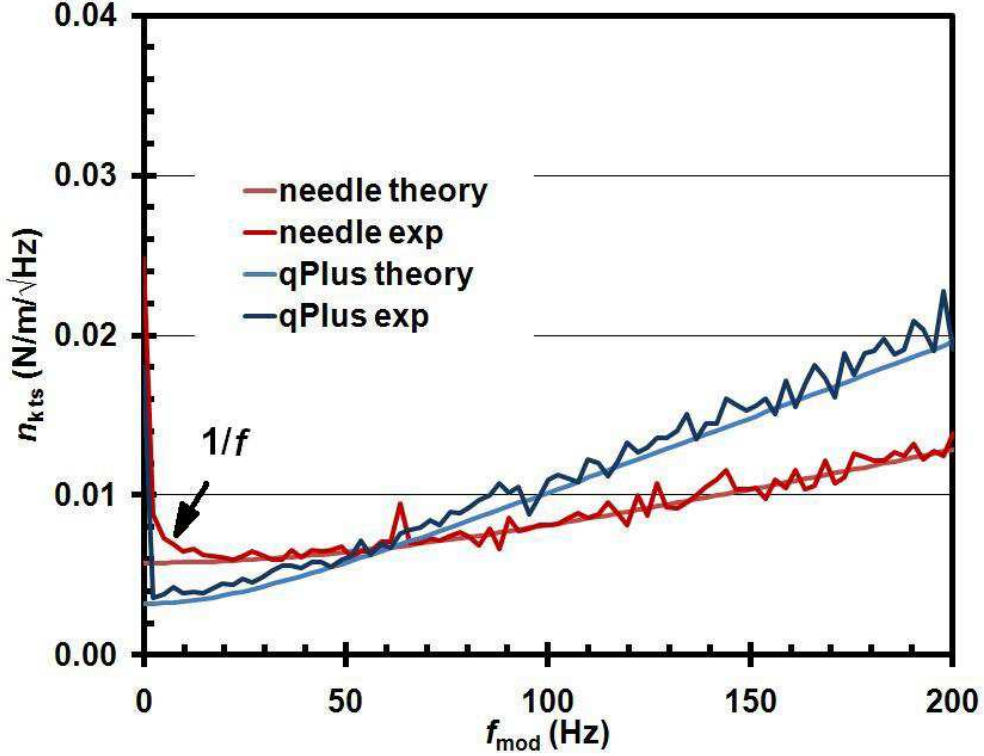}
\end{center}
\caption{(Color online) Total experimental and calculated
force-gradient-noise-densities as a function of modulation
frequency for the needle sensor (red line) and qPlus sensor (black
line) at room temperature. The calculated force
force-gradient-noise-densities are derived with the experimental
values for $S, k, n_{amp}, Q$ and $f_0$ at an amplitude of
$A=100$\,pm. The $1/f$-component for small $f_{mod}$ is due to
thermal frequency drift noise (see eq. \ref{eq_n_kts_drift}).} \label{fig_n_kts_fmod}
\end{figure}

Figure \ref{fig_n_kts_fmod} shows the calculated (smooth lines) and
experimental (jagged lines) power spectral density of the force
gradient noise $n_{kts}$ as a function of modulation frequency $f_{mod}$. This graph is
produced by inserting the output of the phase-locked-loop detector to a FFT Analyzer (Agilent)
and multiplying the frequency shift by the corresponding scaling factor ($k_{ts}=2k/f_0 \times \Delta f$, thus $n_{kts}=2k/f_0 \times n_{\Delta f}$).

All four noise sources contribute to the experimental noise graphs.
The absolute force gradient noise figures outlined in section \ref{sec_noise} can be transformed in a density representation by
\begin{equation}\label{eq_del_kts}
    n_{k\,ts}(f_{mod}) = \sqrt{\frac{\partial \delta k_{ts}^2}{\partial B} |_{B=f_{mod}}}.
\end{equation}
Thus, we can explicitly calculate the four spectral noise contributions from
quantities that can be obtained from the thermal noise spectrum as shown in Fig. \ref{fig_n_vout_qp} and a
measurement of sensor stiffness.
\begin{enumerate}
          \item
          For the detector noise contribution, we find
          \begin{equation}
    n_{k\,ts\,det}(f_{mod}) = \sqrt{8}\frac{k n_{q}}{f_0 A} f_{mod}.
\end{equation}
          \item
          Thermal noise is constant with respect to $f_{mod}$:
\begin{equation}
    n_{k\,ts\,th} = \sqrt{\frac{4k k_{B}T}{\pi A^2 f_0 Q}}.
\end{equation}
          \item Oscillator noise is also constant with $f_{mod}$:
\begin{equation}
    n_{k\,ts\,osc} = \sqrt{2}\frac{k n_{q}}{Q A}.
\end{equation}
          \item Frequency drift noise is inversely proportional to $f_{mod}$:
\begin{equation}
n_{kts\,drift}(f_{mod})=\frac{2 k r\sqrt{\tau}}{f_0 \pi f_{mod}}.
\end{equation}
        \end{enumerate}

The total noise of the force gradient measurement is given by
\begin{equation}\label{eq_del_kts2}
    \delta k_{ts}=\sqrt{\int_{1/\tau}^B n_{kts}^2(f_{mod}) df_{mod}}
\end{equation}
with
\begin{equation}\label{eq_del_kts3}
    n_{kts}^2(f_{mod}) = n_{k\,ts\,det}^2(f_{mod}) + n_{k\,ts\,th}^2 + n_{k\,ts\,osc}^2 + n_{kts\,drift}(f_{mod})^2.
\end{equation}

The calculated graphs include deflection detector noise (linear
with $f_{mod}$), thermal noise (constant with $f_{mod}$) and oscillator noise (also constant with $f_{mod}$).
Frequency drift noise, which is large for long measuring times (i.e. small $f_{mod}$) is not included in the calculation, but clearly apparent in the measurement by
the increase of the experimental needle deflection detector noise density for small $f_{mod}$.
As expected, the qPlus sensor shows less thermal, oscillator and frequency drift noise, but more detector noise. This is due to the excellent adaption of the
FEMTO/Kolibri amplifier \cite{Femto,KolibriPreamp} to the needle sensor and to the fact, that the standard qPlus sensor as described in Table \ref{table_sensordims} only has 50\% of the calculated sensitivity.

Table \ref{table_noise} summarizes the results in a way that all noise contributions can be identified.
\begin{center}
\begin{table}
\begin{tabular}{|c||c|c|c|c|c|c|c|}
    \hline
    sensor     &  $n_q$ & $Q$ & $\frac{\delta k_{ts\,det}}{B^{3/2}}$    & $\frac{\delta k_{ts\,th}}{B^{1/2}}$    & $\frac{\delta k_{ts\,osc}}{B^{1/2}}$    & $\delta k_{ts\,drift}$ (300\,K)  & $\delta k_{ts\,drift}$ (4\,K)   \\
               &  $\frac{\textrm{fm}}{\textrm{Hz}^{1/2}}$& & $\frac{\mu\textrm{N/m}}{\textrm{Hz}^{3/2}}$ &  $\frac{\mu\textrm{N/m}}{\textrm{Hz}^{1/2}}$  & $\frac{\mu\textrm{N/m}}{\textrm{Hz}^{1/2}}$ &   $\frac{\textrm{mN}}{\textrm{m}}$, $\Delta T = 0.1$\,K  &  $\frac{\textrm{mN}}{\textrm{m}}$, $\Delta T = 10$\,mK   \\
    \hline \hline
    qPlus 300\,K air&       62  &   2900  & 60.7               & 3290                    &   544                & 0.05              &         \\
    \hline
    qPlus 300\,K UHV&       62  &   5000 & 60.7               & 2510                    &   316                & 0.05              &         \\
    \hline
    qPlus   4\,K UHV&       31  &  200000 & 30.4               &   46                  &     8                &                   &   0.036 \\
    \hline
    needle 300\,K air&      1.89&   18500 & 33.4               & 5530                    &  1560                & 31                &         \\
    \hline
    needle 300\,K UHV&      1.89&   50000 & 33.4               & 3370                   &  577                & 31                &         \\
    \hline
    needle  4\,K UHV&       1.89&   80000 & 33.4               &  308                    &  361                 &                   &    21.6 \\
    \hline
\end{tabular}
\caption{Noise contributions of the four noise sources for qPlus ($f_0=30$\,kHz)
and needle sensor ($f_0=998$\,kHz for $A=100$\,pm and $B=1$\,Hz. Note that
detector noise scales with $B^{3/2}$ (after eq. \ref{df_sensor}),
while thermal noise (after eq. \ref{delta_kts_thermal}) and
oscillator noise (after eq. \ref{delta_kts_osc}) scales with
$B^{1/2}$. Thus for $B=100$\, Hz, detector noise would increase by a factor of 1000, while thermal and oscillator noise would only increase by a factor of 10. Frequency drift noise (after eq.
\ref{eq_temp_drift_kts}) is independent of amplitude and becomes
large for small bandwidths. For both sensors, the $\delta
k_{ts\,drift}$ data at 300\,K are based on the parabolic frequency
drift according to Eq. \ref{eq_temp_drift_f} for $T=T_p \pm 2$\,K
while the data at 4\,K are based on a relative frequency drift of
1\,ppm/K (see fig. 2 in \cite{Rychen2000}).}\label{table_noise}
\end{table}
\end{center}

\section{Suggestions for improvements on qPlus and LER sensors}
\label{section_suggestions}
\subsection{Decreasing deflection detector noise}
\begin{figure}
\begin{center}
\includegraphics[clip=true, width=0.8\textwidth]{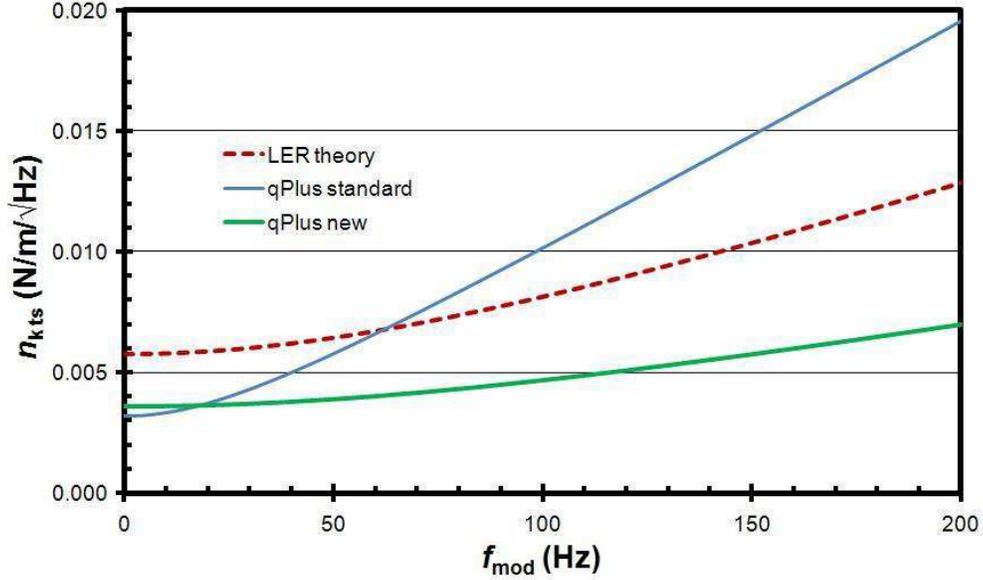}
\end{center}
\caption{(Color online) Calculated force-gradient-noise-densities $n_{kts}$ as a
function of modulation frequency for the standard needle sensor
(red line), qPlus sensor (blue line) and a modified qPlus sensor
with $f_0=92.8$\,kHz, $k=3500$\,N/m, $Q=1650$ and
$n_q=28$\,fm$/\sqrt{\textrm{Hz}}$ (green line). The calculated
values for $n_{kts}$ are based on measured values of $n_{amp}, S,
k, f_0$ and $Q$. } \label{fig_n_kts_newqP}
\end{figure}

With the equations that link signal and noise to the physical
parameters of the sensors, we can now attempt to tailor the design
values for optimal performance. Equation \ref{dkts_qPlus} connects
the relative frequency noise (detector contribution) to the
sensitivity of the sensor and the noise performance of the
amplifier.
For both sensors, we find
\begin{equation}
\delta k_{ts\,det} = 2k\sqrt{ \frac{2}{3}}
\,\frac{n_{amp}}{SAf_0}B^{3/2}.
\end{equation}
With eqs. \ref{LER_k}, \ref{LER_f0} and \ref{eq_S_needlesens_k} in
the ideal situation of $L_e=L$, we can express the spring constant
$k$, sensitivity $S$ and eigenfrequency $f_0$ in terms of the
geometrical parameters $t,w$ and $L$:
\begin{equation}
\delta k_{ts\,det\,ns} = 8 \sqrt{ \frac{2}{3}}
\frac{n_{amp}t}{d_{21}A v_s} B^{3/2}.
\end{equation}
For the qPlus sensor, we use eqs. \ref{k_cl}, \ref{f0_cl} and
\ref{eq_S_qPlus_k} assuming again $L_e=L$, finding
\begin{equation}
\delta k_{ts\,det\,qP} =2.06 \sqrt{ \frac{2}{3}}
\frac{n_{amp}t}{d_{21}A v_s} B^{3/2}.
\end{equation}
This result seems quite surprising: deflection detector noise only
depends on the thickness $t$ of the sensor - all the other
geometrical dimensions cancel, and when comparing a qPlus and a
needle sensor with the same thickness, the qPlus sensor should
only display about 1/4 of the noise of the needle sensor if the
charge noise of the amplifier in use is similar. If we take into
account, that the quartz-cantilever geometry only produces about
50\,\% of the theoretical sensitivity, a qPlus sensor with the
same thickness of a needle sensor should display only 1/2 the
noise. Miniaturisation therefore appears to be the road to
success. The reason for the superior signal-to-noise ratio of the
cantilever geometry implemented in the qPlus sensor over the
length extensional principle utilized in the needle sensor lies in
the fact that the cross section of the qPlus sensor beam shows a
strain and stress profile that is zero in the center and increases
towards the edges, where the charge-collecting electrodes are
located, while the cross section of the needle sensor has a
uniform stress and strain profile (see Fig. \ref{fig_sensorgeom}
d, h). Figure \ref{fig_n_kts_newqP} displays the noise figures of
standard needle and qPlus sensors and a modified qPlus sensor with
a smaller thickness $t$ and smaller length $L$ with
$f_0=92.8$\,kHz, $k=3500$\,N/m, $Q=1650$ and
$n_q=28$\,fm$/\sqrt{\textrm{Hz}}$. This sensor is not only
superior to the needle sensor in thermal, oscillator and frequency
drift noise, but also in detector noise.

\subsection{Decreasing thermal noise}
As outlined in equation \ref{delta_kts_thermal}, the thermal noise
in the force gradient measurement is given by
\begin{equation} \delta
k_{ts\,\, thermal}(z)=\sqrt{\frac{4 k k_{B}TB}{\pi A^{2} f_{0}
Q}},
\end{equation}
Thus, thermal noise
can be minimized by a reduction of temperature, using a stiffness
$k$ as small as possible compatible with stability and choosing a
high eigenfrequency $f_0$ while maintaining a high $Q$-value.

\subsection{Decreasing oscillator noise}
Oscillator noise
can be minimized by combining the recipes to reduce deflection detector noise
and thermal noise, because oscillator noise goes down with decreasing
deflection detector noise, increasing $Q$, and minimizing $k$.

\subsection{Decreasing frequency drift noise}
Again, frequency drift noise is minimized by choosing the appropriate stiffness $k$ of the cantilever.
Because frequency drift noise is proportional to
$k$, we need a stiffness as small as possible (yet allowing stable oscillation at small
amplitudes). A second factor regards temperature stabilization and choosing an operating
temperature close to the turnaround temperature of the corresponding quartz crystal orientation.
Another possibility would be to tailor the turnover temperature
of the quartz crystal by cutting it along the corresponding crystal direction. For the needle
sensor, it might be useful to trigger the frequency detector (PLL) with an atomic clock
because the frequency shift changes can become very small for weakly interacting samples.
More precise measurements on the thermal frequency variation at low temperatures are
needed to assess frequency drift noise for cryogenic microscopes (here, we have
used a value of 1\,ppm/K according to fig. 2 in \cite{Rychen2000}).

\section{Practical considerations regarding tip mounting}
Tip mass plays a crucial role in the needle sensor, because an
imbalance in the effective mass of the coupled beams reduces $Q$.
Rychen has analyzed the effect of mass imbalance and found that
for tuning fork geometries, an imbalance of 1.5\,\% leads to a
drop of the $Q$-value by 63.5\,\% (Fig. 4.8 in \cite{Rychen2001}).
Probably, the effect of mass imbalance is smaller for length
extensional resonators than for tuning forks, however, mass
imbalance will effect the $Q$-value of the needle sensor.
Therefore, the tip of a needle sensor needs to be very small. Long
and thin tips, however, can show significant thermal lateral
oscillations and bend strongly under lateral forces. Youngs
modulus of tungsten is around 400 GPa, thus a wire with a diameter
of 0.01\,mm and a length of 0.3\,mm has a lateral stiffness of
only 22\,N/m. In contrast, the qPlus sensor can easily accommodate
heavy and more stable tips that can be resharpened more easily,
with significant abrasion \cite{Hofmann2010} and even cleaved in
situ \cite{Wutscher2011}.

\section{Noise comparison between large-amplitude (Si cantilevers) and
small amplitude (quartz sensors) operation}
This manuscript focusses on quartz force sensors, but many impressive results have
been obtained with AFM using Si cantilevers, such as high-resolution force spectroscopy \cite{Lantz2001}
imaging the rest atoms on Si(111)-(7$\times$7) \cite{Lantz2000,Eguchi2002}, imaging of insulators
\cite{Barth2001,Reichling2001}, atomic manipulation \cite{Sugimoto2005}, chemical identification \cite{Sugimoto2007} and the detection of short range magnetic exchange forces \cite{Kaiser2007}.
It is instructive to compare
the noise performance of quartz sensors with silicon cantilevers. When comparing
only the thermal force gradient noise for silicon cantilevers and
quartz sensors (see Table I in \cite{Torbrugge2010}), Si
cantilevers appear to be superior by more than four orders of magnitude. However, we
need to consider that Si cantilevers can not be operated in the
force gradient regime when the tip comes close enough to feel
chemical bonding forces \cite{Giessibl1997,Giessibl2004N}. Standard Si
cantilevers need to be operated at amplitudes of a few ten
nanometers, and the frequency shift is in that case given by the
normalized frequency shift $\gamma$ \cite{Giessibl1997} with
\begin{equation}
\gamma = \frac{\Delta f}{f_{0}}k A^{3/2}\approx
\frac{1}{\sqrt{2\pi}}F_{ts}\lambda^{1/2},
\end{equation}
where $F_{ts}$ is the tip-sample force and $\lambda$ is its range
\cite{Giessibl1997,Giessibl2000}. For small amplitude operation,
we find
\begin{equation}
k_{ts} = 2 k \frac{\Delta f}{f_{0}}.
\end{equation}
While we cannot compare a minimal detectable force gradient and a
minimal detectable normalized frequency shift, we can calculate a
minimal detectable force $\delta F_{ts\,min}$ for a given range
$\lambda$. For the large amplitude regime, we find
\begin{equation}
\delta F_{ts\,min} = \sqrt{2\pi}k \frac{\delta  \Delta
f_{min}}{f_{0}} \frac{A^{3/2}}{\lambda^{1/2}}.
\end{equation}
For small amplitudes, the force noise is given by the product
between the minimal detectable force gradient and the range:
\begin{equation}
\delta F_{ts\,min} = 2k \frac{\delta  \Delta
f_{min}}{f_{0}}\lambda.
\end{equation}
\begin{center}
\begin{table}
\begin{tabular}{|c||c|c|c|c|c|c|c|c|c|}
    \hline
    sensor     &$k$    & $f_0$ & $n_q$                                  & $Q$ &$A$&  $\frac{\delta F_{ts\,det}}{B^{3/2}}$    & $\frac{\delta F_{ts\,th}}{B^{1/2}}$    & $\frac{\delta F_{ts\,osc}}{B^{1/2}}$    & $\delta F_{ts\,drift}$ (300\,K)      \\
               & N/m & kHz  & $\frac{\textrm{fm}}{\textrm{Hz}^{1/2}}$&   &nm   & $\frac{\textrm{fN}}{\textrm{Hz}^{3/2}}$ &  $\frac{\textrm{fN}}{\textrm{Hz}^{1/2}}$  & $\frac{\textrm{fN}}{\textrm{Hz}^{1/2}}$ &   pN, $\Delta T = 0.1$\,K      \\
    \hline \hline
 Si cantilever \cite{Eguchi2002},\cite{Torbruegge2008} &46     & 298.0& 272                                    &54200 &  4  &0.6                                    & 34                                     &  2.9                      & 11.5                          \\
 \hline
 Si cantilever \cite{Fukuma2005} &42     & 281.5& 17                                    &50000 & 8  &0.04                                    & 5                                     &  0.3                      & 29.7                          \\
 \hline
    qPlus opt. det. \cite{Morita2010}      & 1500  & 27.8 &  15                                    &6100 & 0.1 & 1.0                                     & 170                                    &  4.1                      & 0.003                        \\
    \hline
    qPlus el. det.     & 1800  & 32.8 &  62                                    &2900 & 0.1 & 4.8                                     & 260                                    &  43                      & 0.004                        \\
    \hline
    needle     &1080000&1000&   1.89                                 &18500 & 0.1 & 2.6                                     & 437                                    &  123                     & 2.5                          \\
    \hline

\end{tabular}
\caption{Noise contributions of the four noise sources for different Si
cantilevers, qPlus and needle sensor and
$B=1$\,Hz with respect to an exponential attractive force with
$\lambda=79$\,pm (Morse potential, as shown in Fig.
\ref{fig_deltafMorse}).}\label{table_SiCl}
\end{table}
\end{center}
As shown in table \ref{table_SiCl}, Si cantilevers with refined optical readout schemes are better in detector, thermal and oscillator noise but show profoundly larger thermal drift noise. Also shown are the calculated noise figures for a qPlus sensor with optical deflection detection, reaching lower values for detector noise than in the electrically detected mode \cite{Morita2010}.
At low temperatures, the first three noise types decrease significantly for quartz sensors, but it is unclear how effective a Si cantilever can be cooled even in a low temperature environment
when intense laser light from the optical deflection detector is shined on them. Although Si cantilevers with good optical deflection detectors show less noise than quartz cantilevers, detector noise, thermal noise and oscillator noise can be reduced by bandwidth reduction, and the thermal drift noise is significantly smaller for quartz cantilevers than for Si cantilevers.

\section{Summary and Outlook}

Concluding, we compared force sensors based on length-extensional
resonators and based on quartz tuning forks.
We found that in contrast to applications in the literature, the effective spring
constant of a needle sensor is actually twice as large as the stiffness of one tine (see eq. \ref{eq_keff}).
We have discussed
four types of noise: deflection detector noise, thermal noise, oscillator noise
and frequency drift noise. Surprisingly, the deflection detector noise is
independent of sensor stiffness, because while a stiffer sensor
has less frequency shift proportional to $1/k$, its deflection
signal increases linear with $k$. The other three noise sources, however,
clearly favor sensors with spring constants around 1\,kN/m. The cantilever geometry provides
more charge per force than the length extensional geometry. However, the
longitudinal outline of the needle sensor is more suited to a space conserving
microscope.

\section{Acknowledgments}
We thank Federico de Faria-Elsner, Joachim Welker and Jay Weymouth
for writing the software that analyzes the thermal peak and
deflection noise densities. We are grateful to Stefan Torbr\ügge,
Gerhard Meyer and Fabian Mohn for helpful comments.

\end{document}